\documentclass[fleqn,usenatbib,usedcolumn]{mnras}

\usepackage[T1]{fontenc}

\DeclareRobustCommand{\VAN}[3]{#2}
\let\VANthebibliography\thebibliography
\def\thebibliography{\DeclareRobustCommand{\VAN}[3]{##3}\VANthebibliography}

\usepackage[british]{babel}
\usepackage{graphicx}	
\usepackage{amsmath}	
\usepackage{amssymb}	

\usepackage{newtxtext}
\usepackage[slantedGreek]{newtxmath}

\usepackage[inline]{enumitem}
\usepackage{soul} 
\usepackage{amsmath}
\usepackage{newtxtext}
\usepackage[slantedGreek]{newtxmath}
\usepackage{xifthen}
\usepackage{xspace}

\usepackage[dvipsnames]{xcolor}
\usepackage{ulem}
\definecolor{harvestgold}{rgb}{0.85, 0.57, 0.0}
\definecolor{darkblue}{RGB}{0, 0, 139}

\newcommand{\orcid}[1]{\href{https://orcid.org/#1}{\,\includegraphics[height=\fontcharht\font`\B]{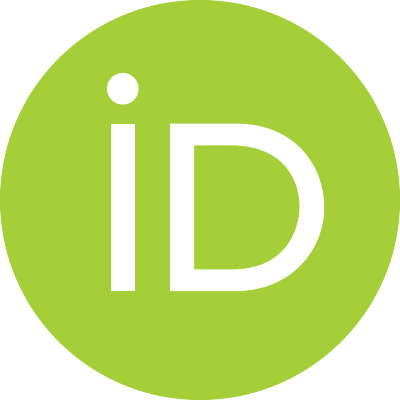}}}
\newcommand{\code}[1]{\texttt{\detokenize{#1}}}
\newcommand{\percent}[1]{\ensuremath{#1}~per~cent}
\newcommand{\hermeian}{Hermeian}

\DeclareMathOperator{\sign}{sgn}
\newcommand{\bigO}[1]{\ensuremath{\mathop{}\mathopen{}\mathcal{O}\mathopen{}\left(#1\right)}}

\newcommand{\PLANCKthirteenp}{\ensuremath{{\Ho{}=\kms[67.7]\pMpc{}},\,\allowbreak {\Omegaup_{\rm M}=0.318},\,\allowbreak {\Omegaup_\Lambdaup=0.682},\,\allowbreak {\Omegaup_{\rm b}=0.048},\,\allowbreak {n_s=0.968},\,\allowbreak {\sigma_8=0.83}}}

\newcommand{\appref}[1]{Appendix~\ref{#1}}
\newcommand{\secref}[1]{Section~\ref{#1}}

\newcommand{\figref}[1]{Fig.~\ref{#1}}
\newcommand{\figrefs}[2]{Figs~\ref{#1}~and~\ref{#2}}
\newcommand{\eqnref}[1]{eq.~(\ref{#1})}

\newcommand{\exteqns}[2]{eqs~(\ensuremath{#1})~and~(\ensuremath{#2})}

\newcommand{\extsec}[1]{section~\ensuremath{#1}}

\newcommand{\Fermi}{{\textit{Fermi}--LAT}}

\newcommand{\Arepo}{{\sc AREPO}}
\newcommand{\Auriga}{{\sc AURIGA}}
\newcommand{\Hestia}{{\sc HESTIA}}

\newcommand{\AHF}{{\sc ahf}}
\newcommand{\astropy}{{\sc Astropy}}
\newcommand{\MergerTree}{{\sc MergerTree}}
\newcommand{\numpy}{{\sc numpy}}
\newcommand{\python}{{\sc python}}
\newcommand{\scipy}{{\sc scipy}}
\newcommand{\matplotlib}{{\sc matplotlib}}

\newcommand{\LCDM}{{$\Lambdaup$CDM}}
\newcommand{\conc}[1]{\ensuremath{c_{#1}}}
\newcommand{\cvir}{\ensuremath{\conc{200}}}
\newcommand{\Jfactor}{\textit{J}-factor}
\newcommand{\M}[1]{\ensuremath{M_{#1}}}
\newcommand{\Mvir}{\ensuremath{\M{200}}}
\newcommand{\R}[1]{\ensuremath{R_{#1}}}
\newcommand{\Rvir}{\ensuremath{\R{200}}}

\newcommand{\PLANCK}[1]{{PLANCK~\ensuremath{#1}}}

\newcommand{\unit}[1]{\ensuremath{\mathrm{\,#1}}\xspace}
\newcommand{\unitlogicspace}[2]{%
  \ifthenelse{\isempty{#1}}%
    {\unit{#2}}
    {\ensuremath{{{#1}\, \unit{#2}}}}
  }

\newcommand{\Msun}[1][]{\unitlogicspace{#1}{M_\odot}}
\newcommand{\Mpc}[1][]{\unitlogicspace{#1}{Mpc}}
\newcommand{\kpc}[1][]{\unitlogicspace{#1}{kpc}}
\newcommand{\pMpc}[1][]{\unitlogicspace{#1}{Mpc^{-1}}}
\newcommand{\GeV}[1][]{\unitlogicspace{#1}{GeV}}
\newcommand{\TeV}[1][]{\unitlogicspace{#1}{TeV}}
\newcommand{\Gyr}[1][]{\unitlogicspace{#1}{Gyr}}
\newcommand{\Myr}[1][]{\unitlogicspace{#1}{Myr}}
\newcommand{\kms}[1][]{\unitlogicspace{#1}{km\, s^{-1}}}

\newcommand{\variablelogicspace}[2]{%
  \ifthenelse{\isempty{#2}}%
    {\ensuremath{#1}}
    {\ensuremath{{{#1}={#2}}}}
  }
\newcommand{\Ho}[1][]{\variablelogicspace{H_0}{#1}}
\newcommand{\Nperi}[1][]{\variablelogicspace{N_{\rm peri}}{#1}}
\newcommand{\thetamid}[1][]{\variablelogicspace{\theta_{\rm mid}}{#1}}
\newcommand{\tlb}[1][]{\variablelogicspace{t_{\rm lookback}}{#1}}
\newcommand{\rmid}[1][]{\variablelogicspace{r_{\rm mid}}{#1}}
\newcommand{\Vmax}[1][]{\variablelogicspace{V_{\rm max}}{#1}}
\newcommand{\z}[1][]{\variablelogicspace{z}{#1}}
\graphicspath{{Figures/}}

\title[Hermeian haloes in the Local Group]{\hermeian{} haloes: Field haloes that interacted with both the Milky Way and M31}

\author[O. Newton et al.]{Oliver Newton\orcid{0000-0002-2769-9507}\thanks{E-mail: o.j.newton@ljmu.ac.uk},$^{1,2}$
Noam~I.~Libeskind\orcid{0000-0002-6406-0016},$^{3,1}$
Alexander Knebe,$^{4,5,6}$
Miguel~A.~S\'{a}nchez-Conde\orcid{0000-0002-3849-9164},$^{7}$
\newauthor
Jenny~G.~Sorce,$^{8,3}$
Sergey Pilipenko,$^{9}$
Matthias Steinmetz\orcid{0000-0001-6516-7459},$^{3}$
R\"{u}diger Pakmor\orcid{0000-0003-3308-2420},$^{10}$
Elmo Tempel\orcid{0000-0002-5249-7018},$^{11,12}$
\newauthor
Yehuda Hoffman,$^{13}$
Mark Vogelsberger\orcid{0000-0001-8593-7692}$^{14}$
\\
$^{1}$Univ. Lyon, Univ. Claude Bernard Lyon 1, CNRS, IP2I Lyon/IN2P3, IMR 5822, F-69622, Villeurbanne, France\\
$^{2}$Astrophysics Research Institute, Liverpool John Moores University, 146 Brownlow Hill, Liverpool L3 5RF, UK\\
$^{3}$Leibniz-Institut f\"{u}r Astrophysik Potsdam (AIP), An der Sternwarte 16, D-14482 Potsdam, Germany\\
$^{4}$Departamento de F\'{i}sica Te\'{o}rica, M\'{o}dulo 8, Facultad de Ciencias, Universidad Aut\'{o}noma de Madrid, 28049 Madrid, Spain\\
$^{5}$Centro de Investigaci\'{o}n Avanzada en F\'{i}sica Fundamental (CIAFF), Facultad de Ciencias, Universidad Aut\'{o}noma de Madrid, 28049 Madrid, Spain\\
$^{6}$International Centre for Radio Astronomy Research, University of Western Australia, 35 Stirling Highway, Crawley, Western Australia 6009, Australia\\
$^{7}$Institute for Theoretical Physics (IFT UAM/CSIC) and the Department of Theoretical Physics, Universidad Aut\'{o}noma de Madrid, 28049 Madrid, Spain\\
$^{8}$Univ. Lyon, ENS de Lyon, Univ. Lyon~1, CNRS, Centre de Recherche Astrophysique de Lyon, UMR5574, F-69007, Lyon, France\\
$^{9}$P.N. Lebedev Physical Institute of the Russian Academy of Sciences, Profsojuznaja 84/32 Moscow, Russia, 117997\\
$^{10}$Max-Planck-Institut f\"{u}r Astrophysik, Karl-Schwarzschild-Str. 1, D-85748, Garching, Germany\\
$^{11}$Tartu Observatory, University of Tartu, Observatooriumi 1, 61602 T\~{o}ravere, Estonia\\
$^{12}$Estonian Academy of Sciences, 10130 Kohtu 6, Tallinn, Estonia\\
$^{13}$Racah Institute of Physics, Hebrew University, Jerusalem, 91904 Israel\\
$^{14}$Department of Physics, Massachusetts Institute of Technology, 77 Massachusetts Avenue, Cambridge, MA 02139, USA
}

\date{Accepted 2022 April 27. Received 2022 April 01; in original form 2021 April 26}

\pubyear{2022}

\begin{document}
\label{firstpage}
\pagerange{\pageref{firstpage}--\pageref{lastpage}}
\maketitle

\begin{abstract}
The Local Group is a unique environment in which to study the astrophysics of galaxy formation. The proximity of the Milky Way and M31 enhances the frequency of interactions of the low-mass halo population with more massive dark matter haloes, which increases their concentrations and strips them of gas and other material. Some low-mass haloes pass through the haloes of the Milky Way or M31 and are either ejected into the field or exchanged between the two primary hosts. We use high resolution gas-dynamical simulations to describe a new class of field haloes that passed through the haloes of both the Milky Way \textit{and} M31 at early times and are almost twice as concentrated as field haloes that do not interact with the primary pair. These `\hermeian{}' haloes are distributed anisotropically at larger distances from the Local Group barycentre than the primary haloes and appear to cluster along the line connecting the Milky Way and M31. Hermeian haloes facilitate the exchange of dark matter, gas, and stars between the Milky Way and M31 and can enhance the star formation rate of the gas in the primary haloes during their interactions with them. We also show that some \hermeian{} haloes can host galaxies that, because they are embedded in haloes that are more concentrated than regular field haloes, are promising targets for indirect dark matter searches beyond the Milky Way virial radius and can produce signals that are competitive with those of some dwarf galaxies. \hermeian{} galaxies in the Local Group should be detectable by forthcoming wide-field imaging surveys.
\end{abstract}

\begin{keywords}
Local Group -- Galaxy: evolution -- galaxies: dwarf -- galaxies: evolution -- galaxies: interactions -- dark matter
\end{keywords}



\section{Introduction}
\label{sec:Introduction}
The Local Group environment is an exceptional probe of fundamental theories, and models of galaxy evolution. The proximity of its constituent galaxies facilitates detailed observations that enable the comprehensive exploration of the physics of galaxy formation across several orders of magnitude in mass. Its population of faint galaxies has also proved to be a compelling test of cosmological models on small astrophysical scales, ruling out various dark matter~(DM) candidates \citep[e.g.][]{lovell_satellite_2016,enzi_joint_2021,nadler_constraints_2021,newton_constraints_2021} and revealing possible small-scale challenges to the prevailing paradigm \citep[see][for recent reviews]{bullock_small-scale_2017,pawlowski_planes_2018}.

Over the last decade, self-consistent Local Group-like volumes have been modelled in detail using increasingly sophisticated simulations \citep[e.g.][]{gottlober_constrained_2010,libeskind_constrained_2010,garrison-kimmel_elvis:_2014,yepes_dark_2014,carlesi_constrained_2016,sawala_apostle_2016,fattahi_apostle_2016,libeskind_hestia_2020}. They have revealed a dynamic environment that influences the growth of the Milky Way~(MW) and Andromeda~(M31) and has important effects on the evolution of the low-mass galaxy population \citep{benitez-llambay_dwarf_2013}. In particular, the simulations show that low-mass objects often interact with more massive haloes in the dense environment of the Local Group. The tidal interactions they experience increase the concentrations of their DM haloes \citep{li_assembly_2013,bakels_pre-processing_2021} and baryonic processes efficiently strip gas from the galaxies they host.

Such `pre-processed' haloes are commonly found in the vicinity of the two `primary' hosts, which dominate the Local Group's internal dynamics. Most are `backsplash' haloes that fell into the halo of the MW or M31 at earlier times and passed back into the field, reaching distances many times the virial radius of the host by the present day \citep{gill_evolution_2005,moore_age-radius_2004,knebe_luminosities_2011}. Backsplash haloes compose more than \percent{10} of field haloes in the Local Group and could account for as many as half of all systems accreted by the MW and M31 \citep{ludlow_unorthodox_2009,teyssier_identifying_2012,garrison-kimmel_elvis:_2014,bakels_pre-processing_2021,green_tidal_2021}. Galaxies in backsplash haloes typically have higher quenched fractions and lower mass-to-light ratios and gas fractions than isolated field galaxies that have not experienced such halo--primary halo interactions \citep{knebe_luminosities_2011,simpson_quenching_2018,buck_nihao_2019}. There is evidence of these processes in some nearby dwarf galaxies, which could have passed close to the MW \citep[e.g.][]{besla_are_2007,pawlowski_perseus_2014,buck_nihao_2019,blana_dwarfs_2020,mcconnachie_solo_2021,putman_gas_2021}. The proximity of the MW and M31 during the assembly of the Local Group also facilitates the exchange of DM haloes between them. These `renegade' haloes pass through one of the primary haloes and are then accreted into the other, potentially transferring baryonic material in the process \citep{knebe_renegade_2011}. As many as half of the satellite galaxies in the MW and M31 could have experienced such interactions \citep{wetzel_satellite_2015}. Renegade haloes share many properties with other pre-processed haloes but differ in their spatial distribution, which is anisotropic and points towards the two host haloes. This is a result of their trajectory through the primary haloes of the Local Group \citep{libeskind_preferred_2011}.

Populations of highly concentrated DM haloes could be competitive probes with which to constrain the properties of the DM. Candidate particles that decay or annihilate with each other in regions of high DM density are predicted to emit electromagnetic radiation, typically in X-rays or gamma-rays. For annihilating DM, the luminosity of the radiation originating from these particle interactions scales strongly with halo mass and concentration \citep{sanchez-conde_flattening_2014}. Such signals have been claimed to be observed already in DM-dominated systems such as the Galactic centre \citep[the so-called `\GeV{} excess';][]{goodenough_possible_2009,calore_tale_2015,daylan_characterization_2016,ackermann_fermi_2017}, the centre of M31 \citep{ackermann_observations_2017}, and the Reticulum~II dwarf galaxy \citep{geringer-sameth_indication_2015}. However, an astrophysical origin of the signal is also possible and has not been ruled out \citep{carlson_cosmic_2014,petrovic_galactic_2014,bartels_strong_2016,lee_evidence_2016,di_mauro_multimessenger_2021}. Gas-poor, pre-processed dwarf galaxies could therefore be attractive targets to break the degeneracy in the origins of the gamma-ray excesses.

In this paper, we describe a new class of concentrated DM field halo, which is a promising target for indirect DM searches. These `\hermeian{}'\footnote{In Ancient Greek mythology, Hermes was the messenger of the gods and the protector of travellers and wayfarers. He also presided over the crossing of thresholds and boundaries and was often invoked alongside the goddess Hestia during times of change and transition.} haloes interact with the haloes of both the MW {\it and} M31 during the assembly of the Local Group and pass back into the field by \z[0]. We study these using high-resolution magnetohydrodynamic simulations of Local Group volumes from the \Hestia{} suite \citep{libeskind_hestia_2020} that we introduce in \secref{sec:Methods}. These simulations are constrained by observations of the peculiar velocity field to reproduce realistic Local Group environments. In \secref{sec:Results}, we investigate how interactions with the primary haloes affect the evolution of the \hermeian{} haloes compared with the backsplash sub-population and the rest of the field haloes and describe their present-day properties by which they might be identified in observations. We also evaluate how \hermeian{} haloes would appear in observational searches for DM annihilation signals. We discuss these results and present concluding remarks in \secref{sec:Disc_Conc}.

\section{Methodology}
\label{sec:Methods}
The \Hestia{} suite consists of 13 medium- and three high-resolution \LCDM{} magnetohydrodynamic cosmological simulations of the Local Group. The initial conditions are constrained by observations of the peculiar velocity field \citep[catalogued by the CosmicFlows-2 survey,][]{tully_cosmicflows-2_2013} to reproduce the major gravitational sources in the local environment. This ensures that at \z[0] the Local Group analogues are embedded in a large-scale cosmography that is consistent with observations when assuming the \LCDM{} cosmological model \citep[see e.g.][]{hoffman_constrained_1991,doumler_reconstructing_2013,sorce_cosmicflows_2016}. Objects such as the Virgo Cluster, the local filament, the supergalactic plane, and the local void are a natural outcome of the constrained simulations. The \Hestia{} simulations are based on the \Arepo{} moving mesh code \citep{springel_e_2010,pakmor_improving_2016,weinberger_arepo_2020} and use the \Auriga{} galaxy formation model that includes a variety of astrophysical processes such as a model for both cold and hot gas in star-forming regions, the exchange of mass and metal content during stellar evolution, and dissipative hydrodynamics \citep{grand_auriga_2017}.

\begin{figure*}%
    \centering%
	\includegraphics[width=\textwidth]{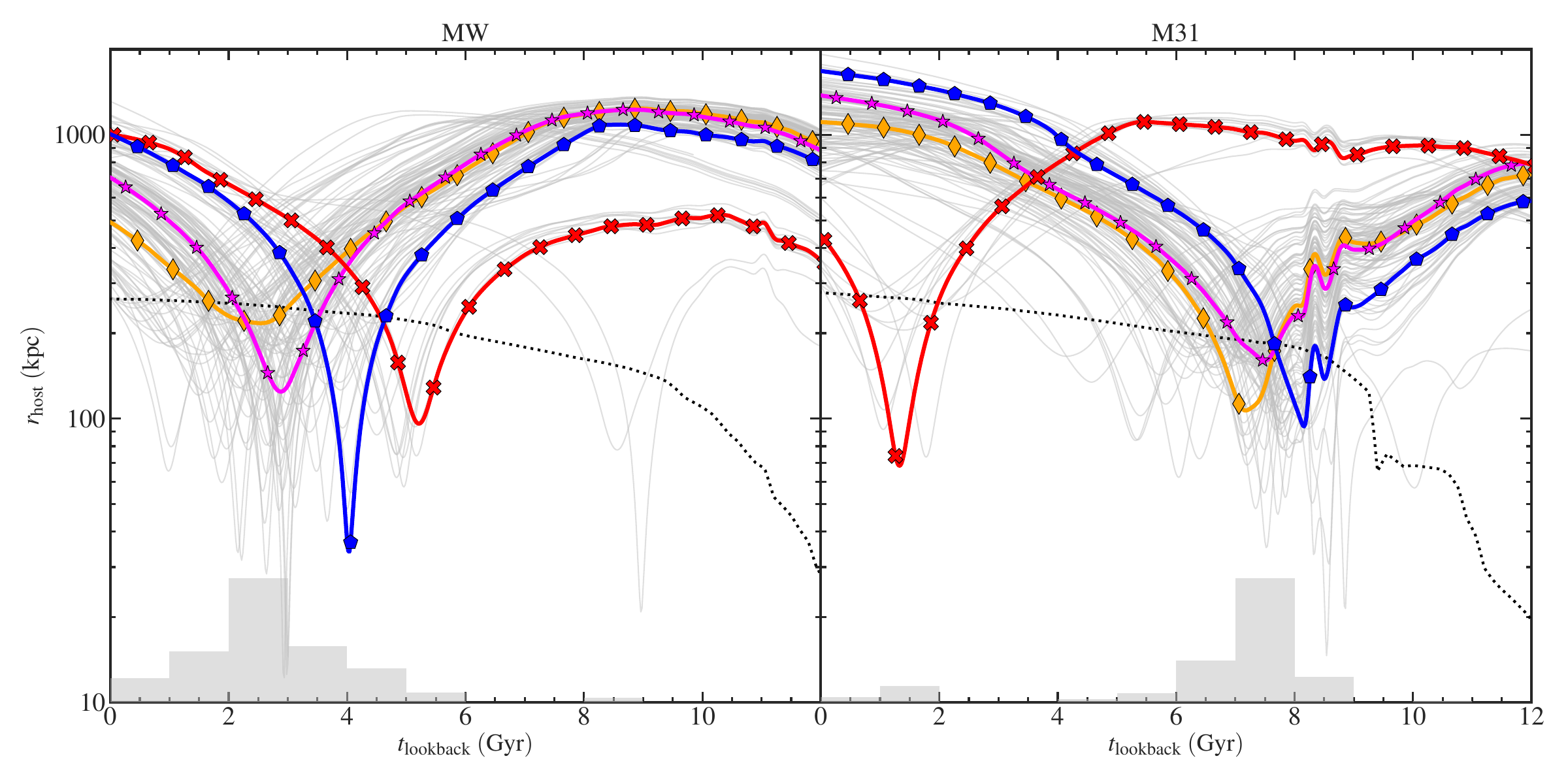}%
	\caption{The distances, $r_{\rm host}$, of the \hermeian{} haloes from the MW and M31 analogues (left-hand and right-hand panels, respectively) as a function of lookback time, $t_{\rm lookback}$. In each panel, the virial radius of the analogue is represented by a dotted line and the trajectories of the `dark' \hermeian{} haloes are plotted as thin solid lines. The four thick solid lines (and the corresponding markers) show the trajectories of the four \hermeian{} galaxies. We plot only objects that are at distances further than \Rvir{} from both host haloes at \z[0]. We show normalized histograms of the times of pericentric passage of all \hermeian{} haloes on the bottom axes. In the right-hand panel, the artefacts in the trajectories at	\tlb[{\Gyr[8.5]}] are caused by the mis-identification of the centre of the M31 analogue in the \AHF{} outputs. There are similar artefacts in the MW analogue at \tlb[{\Gyr[11]}]. They do not affect our results.}%
	\label{fig:Methods:hermeian_distance}%
	\vspace{-10pt}%
\end{figure*}%

Each constrained Local Group is contained within a high-resolution zoom region that accurately resolves a population of dwarf galaxies and the two primary host haloes, which are analogues of the MW and M31 haloes. The geometrical and dynamical configurations of the simulated Local Groups at \z[0] (such as the distance between the primary haloes, their masses and the mass ratio, their line-of-sight velocities, and other properties) match the observations well \citep[see][for more details]{libeskind_hestia_2020}. For this study, we use the three Local Groups that were re-simulated at high resolution (labelled \code{09_18}, \code{17_11}, and \code{37_11}, based on their random seed), each of which uses $8192^3$ effective particles in a composite region of two overlapping spherical volumes centred on the primary haloes that total $\Mpc[{\sim}244]^3$ to achieve DM and gas particle mass resolutions of $\M{\rm DM}=\Msun[2.0\times10^5]$ and $\M{\rm gas}=\Msun[2.2\times10^4]$, respectively. The \Hestia{} suite assumes the \PLANCK{2013} cosmological parameters \citep{planck_collaboration_planck_2014}: \PLANCKthirteenp{}.

We identify gravitationally bound structures and their properties in the high-resolution region of the simulations using the Amiga Halo Finder (\AHF{}) algorithm \citep{gill_evolution_2004,knollmann_ahf_2009}, and exclude from the halo catalogues objects containing fewer than 20 gravitationally bound particles. The evolution of haloes near the edges of the high-resolution region can be disrupted by high-mass simulation particles. To clean the \AHF{} outputs of these we remove haloes that have a low-resolution DM particle within \Rvir{}\footnote{This is the radius at which the mean enclosed matter density is ${\rho\!\left(< \Rvir{}\right) = 200 \times \rho_{\rm crit}}$, where $\rho_{\rm crit}$ is the critical density for closure.} of the halo centre at \z[0]. In our nomenclature, field haloes at \z[0] are at distances further than \Rvir{} from both primary hosts and are within \Mpc[2.5] of the Local Group barycentre. Field haloes that passed within \Rvir{} of only one host halo at least once during the assembly of the Local Group are labelled backsplash haloes \citep{gill_evolution_2005,sales_cosmic_2007,ludlow_unorthodox_2009}, and the \hermeian{} haloes are field haloes that passed within \Rvir{} of \textit{both} hosts during Local Group assembly.

We reconstruct their orbital histories from the main progenitors identified by the \MergerTree{} routine \citep{knebe_impact_2010,srisawat_sussing_2013}. To infer their positions between simulation snapshots we use cubic spline interpolation, which performs well close to the outer radius of the host halo but under-predicts the orbital radii within ${\left[0.15-0.3\right]\times\Rvir{}}$ of the host \citep{richings_subhalo_2020}. For our purposes this does not affect the classification of the haloes as we are only concerned with when they cross inside \Rvir{}, which is not in the affected radial range.

As we are using the most highly resolved simulations available in the \Hestia{} suite, we check our results for numerical convergence using the intermediate-resolution counterparts of the \code{17_11}, \code{37_11}, and \code{09_18} simulations, each of which is simulated using $4096^3$ effective particles. We repeat the analysis described above and find that in the intermediate-resolution simulations the mass functions of the \hermeian{} haloes are converged for halo masses, ${M\!\left(<\Rvir{}\right) = \Mvir{}\gtrsim\Msun[10^8]}$. We expect that comparison with a hypothetical higher resolution simulation using $16\,384^3$ effective particles would show that the high-resolution simulations studied here are converged above halo masses of approximately \Msun[{3\times10^7}]. However, as such a high-resolution counterpart does not exist, we simply assert that our results are robust for haloes with $\Mvir{}\geq\Msun[10^8]$.

\section{Results}
\label{sec:Results}
\begin{figure*}%
    \centering%
	\includegraphics[width=\textwidth]{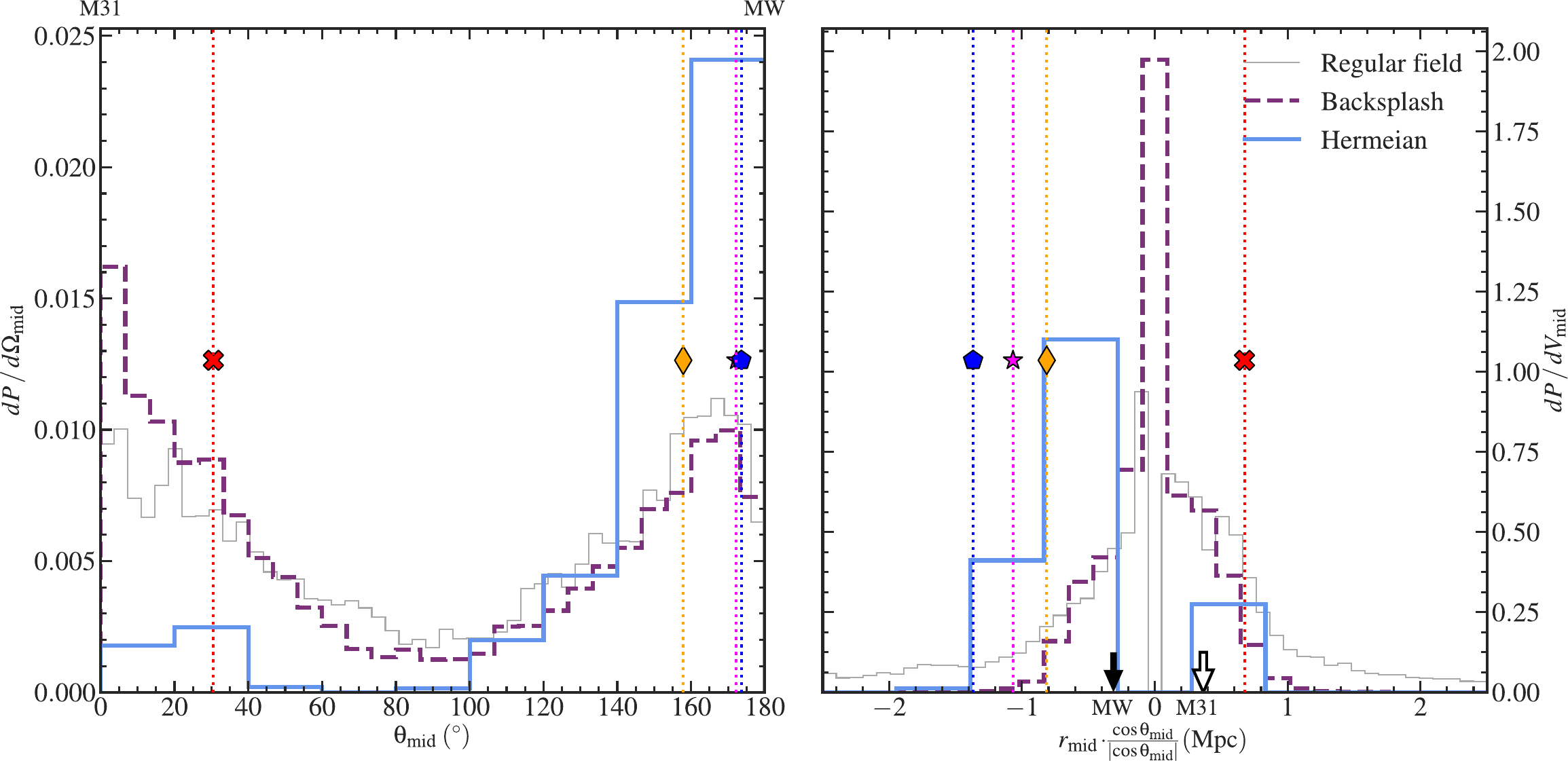}%
	\caption{\textit{Left-hand panel}: Probability distributions of the angles, \thetamid{}, that the backsplash and \hermeian{} sub-populations and the remaining field haloes make with the vector towards M31 from the midpoint of the line connecting the MW and M31 analogues. In this basis, the M31 halo is at \thetamid[{0^\circ}] and the MW is at \thetamid[{180^\circ}].
	\textit{Right-hand panel}: Probability density functions of the distances, \rmid{}, of the haloes from the midpoint of the line. We multiply this by $\sign\left(\cos \thetamid{}\right)$, so that haloes in the hemisphere containing M31 (marked by the unfilled arrow) have positive values and haloes in the other hemisphere containing the MW analogue (marked by the filled arrow) have negative values.
	In both panels, the properties of the \hermeian{} galaxies are indicated by the vertical dotted lines and markers corresponding to the same galaxies as in \figref{fig:Methods:hermeian_distance}.
	}%
	\label{fig:Results:hermeian_midpoint_angle}%
	\vspace{-10pt}%
\end{figure*}%
We find a total of $137$ \hermeian{} haloes in the three high-resolution simulations of the Local Group in \Hestia{}. Over \percent{97} of these do not have any baryonic component, i.e. they are `dark'. However, four contain low-mass galaxies, one of which retains a gaseous component until \z[0]. The four galaxies and $121$ (over \percent{90}) of the total \hermeian{} population are found in the \code{17_11} simulation, so we report results only from this in the analysis that follows. We note that the \code{17_11} simulation is the most massive Local Group in the \Hestia{} suite and its primary haloes are closer together than in the other simulations. The lack of \hermeian{} haloes in the other simulations could be due to the lower masses and larger separations of the primary pairs, although differences in the orientation of the primaries with respect to the Large Scale Structure could also be important.

As we discussed in \secref{sec:Methods}, in the simulations we identify the sub-population of \hermeian{} haloes by studying the past trajectories through the Local Group of present-day field haloes and select those that passed within \Rvir{} of both host haloes. In \figref{fig:Methods:hermeian_distance}, we plot the distance, $r_{\rm host}$, of each \hermeian{} halo from the MW and M31 analogues. The dark \hermeian{} haloes are plotted as thin solid lines and the luminous \hermeian{} galaxies are shown by thick solid lines and distinguishing markers. This shows that at early times the \hermeian{} haloes appear to travel towards one primary halo and away from the other one. This is because the primary haloes are moving away from each other and only move towards each other at later times. We see this behaviour in the other high-resolution simulations in \Hestia{}, although the relative velocities and separations of the primary haloes differ. At \z[0], most \hermeian{} haloes are at distances from the host haloes much larger than \Rvir{}. Adopting a different halo definition, for example using \R{97} that approximates the virial radius in the \PLANCK{2013} cosmology, would likely have little effect on the classification of most \hermeian{} haloes but would remove some that are close to the \Rvir{} boundary at \z[0]. However, it would also expand the classification to include DM haloes that had more shallow, `grazing' interactions with the host, which could affect the distributions of \hermeian{} halo properties \citep[see][for a detailed discussion about the consequences of adopting different definitions of the halo boundary]{diemer_flybys_2021}.

In this simulation, most of the \hermeian{} haloes and three of the \hermeian{} galaxies enter the system as members of a large `association' (i.e. a non-gravitationally bound group) that first fell into the M31 analogue. Using the time of pericentric passage as a metric, there are two main epochs during which the \hermeian{} haloes interact with a host halo. The first is at \tlb[{\Gyr[{8-6}]}], during which \percent{90} of the \hermeian{} haloes in the main group experience pericentric passages with the M31 analogue and are gravitationally attracted towards the MW host halo. Most of the haloes interact with the MW analogue from \tlb[{\Gyr[5]}] until approximately \Gyr[1] before the present day during a second interaction epoch that is \percent{60} longer than the first. This is because the initial interactions with the M31 analogue increased the dispersion of the trajectories of the group members and broadened their angular dispersion relative to the MW host at \z[0]. The latter is shown in \figrefs{fig:Results:hermeian_midpoint_angle}{fig:Results:hermeian_aitoff}, which are discussed below. The remaining dark haloes and the fourth \hermeian{} galaxy, indicated in \figref{fig:Methods:hermeian_distance} by cross symbols, originate from the opposite side of the Local Group and traverse it in the opposite direction, first interacting with the MW analogue and then being attracted towards M31.

There is a limited set of trajectories on which DM haloes can interact with both Local Group primary hosts and move into the field within a Hubble time. This restricts the spatial distribution of \hermeian{} haloes at \z[0]. To characterize this we determine the angle, \thetamid{}, formed by the field haloes and the M31 analogue with respect to the midpoint of the line connecting the two primary hosts. This is an arbitrary choice of basis that sets the M31 analogue at \thetamid[{0^\circ}] and the MW analogue at \thetamid[{180^\circ}]. As we described in \secref{sec:Methods}, we define two sub-categories of field haloes: backsplash haloes, and \hermeian{} haloes. We call the field haloes that have not interacted with either of the two primary haloes `regular' field haloes, to distinguish them from the total field halo population. The backsplash haloes have a similar angular distribution to the regular field haloes in the Local Group; however, a larger proportion of backsplash haloes are close to the MW--M31 line in the direction of the M31 analogue~$\left(\thetamid[{0^\circ}]\right)$ and the \hermeian{} haloes are more strongly concentrated in the direction of the MW analogue~$\left(\thetamid[{180^\circ}]\right)$. These features are related to each other by the large group infall onto the M31 analogue. In addition to supplying most of the \hermeian{} population, many of the DM haloes brought in by the group that did not escape towards the MW stayed close to the M31 host as backsplash haloes.

\begin{figure*}%
    \centering%
	\includegraphics[width=\textwidth]{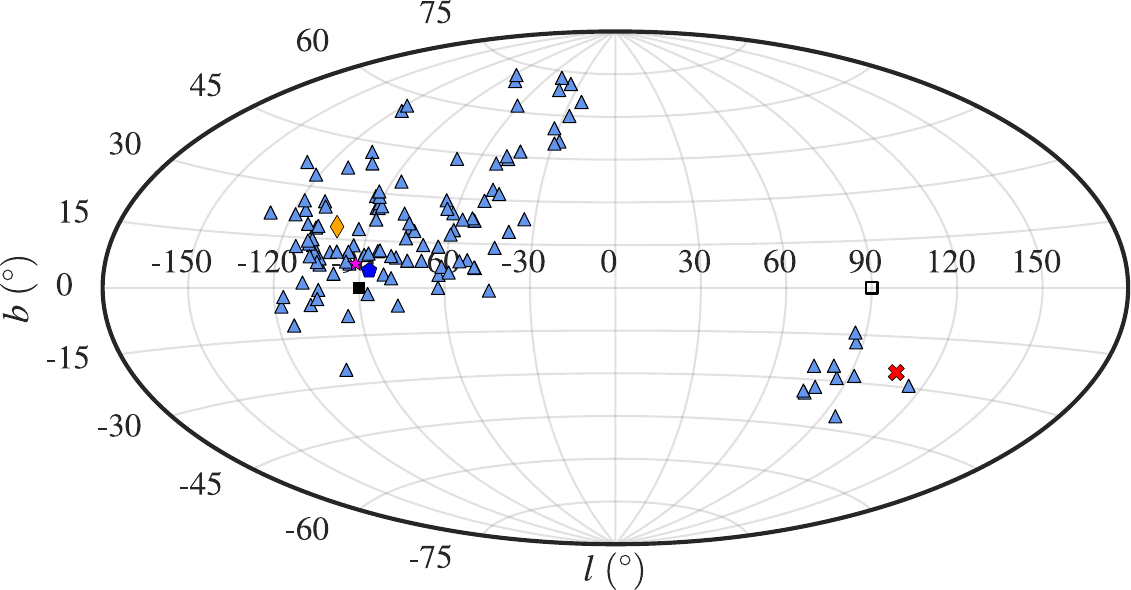}%
	\caption{An Aitoff projection of the \hermeian{} halo population relative to the midpoint of the line connecting the primary pair in the \code{17_11} \Hestia{} simulation. The MW and M31 analogues (at $l=\pm90^\circ$) are marked by filled and unfilled squares, respectively.
	The dark \hermeian{} haloes are plotted with triangles and the four \hermeian{} galaxies are represented by the same markers as in \figref{fig:Methods:hermeian_distance}.}%
	\label{fig:Results:hermeian_aitoff}%
	\vspace{-10pt}%
\end{figure*}%

\begin{figure*}%
    \centering%
	\includegraphics[width=\textwidth]{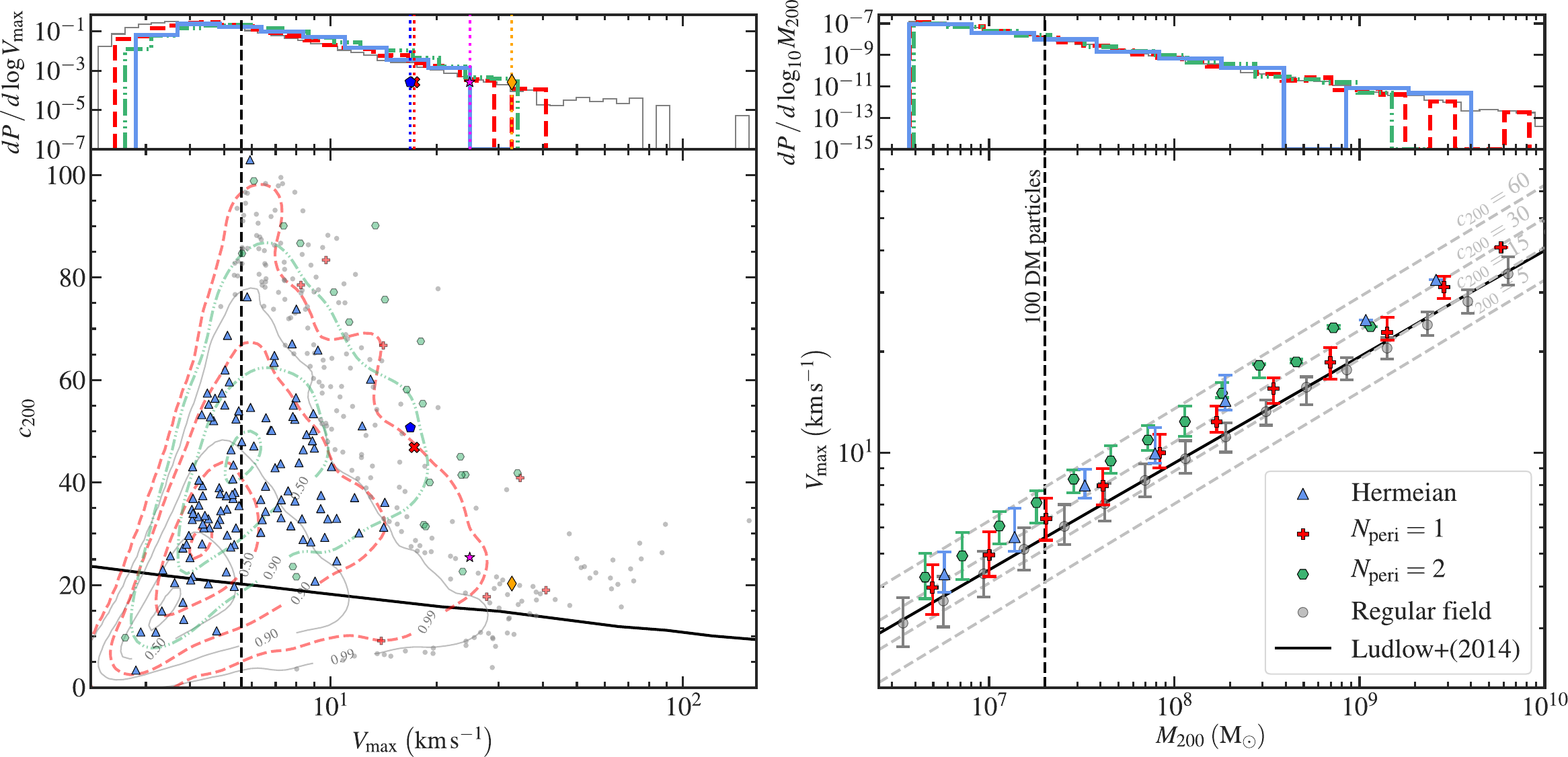}%
	\caption{Properties of the different field halo populations in the \code{17_11} simulation. We label backsplash haloes that experience only one pericentric passage as `\Nperi[1]' and those experiencing two as `\Nperi[2]'. In all panels, we plot the regular field haloes as filled circles and thin solid lines, the \Nperi[1] backsplash haloes as `plus' symbols and thick dashed lines, the \Nperi[2] backsplash haloes as hexagons and thick dot--dot--dash lines, and \hermeian{} haloes as triangles or thick solid lines. We show the four \hermeian{} galaxies using a pentagon, cross, star, and diamond. The vertical dashed lines show the mass of $100$ DM particles in the high-resolution \Hestia{} simulations and the corresponding \Vmax[{\kms[5.59]}], calculated from the median concentration--mass relation of \LCDM{} haloes from \citet{ludlow_massconcentrationredshift_2014}. We show this relationship in both lower panels as solid lines.
	\textit{Upper panels}: Probability distributions of halo maximum circular velocity, \Vmax{}, and total halo mass, \Mvir{}, for each halo population (left and right panels, respectively).
	\textit{Lower left-hand panel}: The concentration, \cvir{}, of haloes as a function of \Vmax{}. The samples of regular field haloes and backsplash haloes are large, so we display them using isodensity contours enclosing $5,\, 50,\, 90$, and \percent{99} of each population (note: we do not plot the \percent{99} contour of the \Nperi[2] population). For each of these populations we plot haloes that are outside the highest isodensity contour. We plot all dark \hermeian{} haloes and the four \hermeian{} galaxies using their respective symbols. Haloes for which \AHF{} could not determine a concentration have not been plotted, at an exclusion rate of less than one~per cent.
	\textit{Lower right-hand panel}: The median \Vmax{}--\Mvir{} relationships of each halo population. The error bars show the \percent{68} scatter in each logarithmically spaced mass bin. The dashed lines show the \Vmax{}--\Mvir{} relationships for NFW DM haloes at different fixed concentrations.
	}%
	\label{fig:Results:c200_vs_vmax_vmax_vs_m200}%
	\vspace{-10pt}%
\end{figure*}%

To strengthen our understanding of the spatial distribution of the Local Group field halo sub-populations, the right-hand panel of \figref{fig:Results:hermeian_midpoint_angle} shows their radial distributions relative to the midpoint of the MW--M31 line. Most of the backsplash and regular field haloes are within \Mpc[1] of the LG midpoint and most commonly are at smaller distances than the two primary haloes \citep[e.g.][]{libeskind_lopsided_2016,pawlowski_lopsidedness_2017,gong_origin_2019,wan_origin_2020}. There are no \hermeian{} haloes within \kpc[400] of the centre of the Local Group and all of them are at larger distances than the MW and M31 analogues (with respect to the LG midpoint); the furthest of these is at a distance of \Mpc[1.6]. Finding a halo at such large distances that has passed within the virial radii of both main haloes without being accreted is particularly surprising. Accounting for the angular distributions of the haloes discussed above, the \hermeian{} population is confined to two lobes at distances further from the Local Group barycentre than the MW and M31. Observational efforts to detect \hermeian{} haloes would be best focused in these areas.

To illustrate the spatial distribution of the \hermeian{} haloes more clearly, in \figref{fig:Results:hermeian_aitoff} we plot the angular positions of the haloes relative to the midpoint of the line connecting the primary pair. Here, the dark \hermeian{} haloes (i.e. those without a baryonic component) are plotted with triangles and the four \hermeian{} galaxies, which retain stars and gas, are plotted with markers corresponding to those used in \figref{fig:Methods:hermeian_distance}. The locations of the MW and M31 analogues are marked with filled and unfilled squares, respectively. As we showed in \figref{fig:Results:hermeian_midpoint_angle}, most of the \hermeian{} haloes lie on the far side of the MW analogue with respect to the Local Group barycentre and close to the line connecting the primary pair. The large angular dispersion is a consequence of the scatter in the trajectories of the group members that was enhanced during their initial interactions with the M31 analogue, lengthening the epoch during which the haloes experienced their pericentric passages with the MW (see \figref{fig:Methods:hermeian_distance}). Nearly \percent{10} of the \hermeian{} haloes closest to the MW analogue are at low galactic latitudes and would likely be obscured by the MW disk, i.e. the Zone of Avoidance. The \hermeian{} haloes closest to the M31 analogue are clustered much closer together in projection, although there are fewer objects in total and they do not all belong to one group.

As the \hermeian{} haloes pass through each Local Group primary, tidal forces suppress further accretion of matter and strip mass from their outer layers \citep[e.g.][]{kravtsov_tumultuous_2004,penarrubia_tidal_2008,warnick_tidal_2008}. Most of this mass loss occurs near pericentre \citep{zavala_dark_2019} concurrently with tidally induced shock heating that reduces the inner density of the infalling halo and increases its concentration \citep{hayashi_structural_2003,kazantzidis_density_2004}. Here, we explore how these dynamical processes affect the properties of the haloes at \z[0]. In the lower left-hand panel of \figref{fig:Results:c200_vs_vmax_vmax_vs_m200}, we compare the concentrations, \cvir{}, of the \hermeian{} and backsplash haloes with the regular field halo population as a function of their maximum circular velocities, \Vmax{}.\footnote{\AHF{} uses a kernel density smoothing technique to calculate \Vmax{} for each halo.} This is a stable proxy for halo mass -- even for objects that experienced some tidal stripping -- that correlates to certain observable properties, such as galaxy rotation curves, more strongly than the halo mass \citep[see discussion in][\extsec{1.2}]{knebe_structure_2013}. \AHF{} calculates halo concentrations under the assumption that the distributions of DM and baryons are described by Navarro--Frenk--White~(NFW) profiles \citep{navarro_simulations_1995,navarro_structure_1996,navarro_universal_1997}.\footnote{\AHF{} calculates \cvir{} using \exteqns{9}{10} of \citet{prada_halo_2012}, i.e. it does not fit a NFW profile but instead uses \Vmax{} and $V_{200}$ to find \cvir{} via an iterative process.} In most cases this is a good choice; however, objects that passed close to one or both of the massive hosts, such as the \hermeian{} and backsplash haloes, have likely been tidally stripped, which truncates their density profiles at large radii \citep{penarrubia_tidal_2008,errani_asymptotic_2021}. For these haloes, we follow the prescription in \citet{errani_asymptotic_2021} and fit the DM distribution using truncated NFW profiles. We compare these to NFW fits to the DM component and find that in almost all cases the truncated profiles are very similar. This is because most of the interacting field haloes escaped the virial radii of the hosts more than \Gyr[0.5] ago, which is equivalent to several tens or hundreds of dynamical times. During this interval the haloes virialize and re-adopt NFW profile forms. Therefore, we use NFW halo concentrations from \AHF{} throughout the rest of this paper except in \secref{sec:Results:Jfactors} where only derived quantities from fits to the DM components of the haloes are relevant.

We categorize the backsplash population according to the number of encounters they have with one of the primaries. We refer to haloes that experienced only one pericentric passage as `\Nperi[1]', and those that experienced two as `\Nperi[2]' backsplash haloes. The interaction history of the latter population is similar to that of the \hermeian{} haloes and their properties evolve in a similar manner. This is not surprising because \hermeian{} haloes can be described as \Nperi[2] backsplash haloes that have had pericentric passages with two different host haloes instead of one. In all cases we exclude haloes for which \AHF{} could not determine a concentration. These account for $0.8,\, 0.6,\, 1.0,$ and \percent{0.3} of the \hermeian{}, \Nperi[1\, {\rm and}\, 2] backsplash, and regular field halo populations, respectively. The haloes that are removed are low mass and all except four regular field haloes have masses smaller than the mass of $100$ DM particles, which we take as an approximate threshold below which the internal structure of haloes is not resolved well in the simulation. In the lower left-hand panel of \figref{fig:Results:c200_vs_vmax_vmax_vs_m200}, at low \Vmax{} there is large scatter in the \cvir{}--\Vmax{} relation of each population; however, the distributions of the data imply that haloes that have interacted with one or both primary hosts (backsplash and \hermeian{} haloes, respectively) are more concentrated than the rest of the field haloes. This is in agreement with previous work examining the backsplash population \citep[e.g.][]{li_assembly_2013,bakels_pre-processing_2021}. The interacting haloes also have higher \Vmax{} than the regular field haloes. Two factors contribute to this:
\begin{enumerate*}
    \item tidal interactions with the host haloes reorganize the internal structure of the interacting haloes, and
    \item numerical effects artificially destroy some low-\cvir{} and low-\Vmax{} haloes in these populations.
\end{enumerate*}
The \Vmax{} functions of the field halo sub-populations, which we discuss in more detail below, show that the enhanced destruction of interacting field haloes because of the limited resolution only affects a narrow range at low \Vmax{}. This does not fully explain the changes in the distribution of haloes in the \cvir{}--\Vmax{} parameter space, and shows that tidal interactions are the dominant effect. The four \hermeian{} galaxies are segregated in \cvir{} according to their mass: the two most massive galaxies (marked by the star and diamond) lie closest to the expected halo mass--concentration relation \citep{ludlow_massconcentrationredshift_2014}, whereas the two lowest mass \hermeian{} galaxies (marked by the cross and pentagon) are much more concentrated at \z[0]. While this is consistent with the effects of tidal interactions that we described above, the evolution of \cvir{} with time is degenerate with the pericentric distances of the \hermeian{} galaxies and could also depend on other properties. As we showed in \figref{fig:Methods:hermeian_distance}, the two most massive \hermeian{} galaxies have the largest pericentric distances, which minimises mass loss during their interactions with the hosts.

In the upper left-hand panel of \figref{fig:Results:c200_vs_vmax_vmax_vs_m200}, we plot the \Vmax{} probability distribution function of each halo population. The \Vmax{} functions of the \hermeian{} and backsplash populations are truncated compared with the regular field halo population. This is caused by tidal interactions that modify their internal structure, remove mass, and reduce their \Vmax{} each time they descend deep into the potential of a Local Group primary. This truncates the upper ends of their \Vmax{} distributions to between \Vmax[{\kms[35-40]}], and the \hermeian{} distribution is affected more severely than the backsplash populations. The \hermeian{} galaxies are hosted by the four \hermeian{} haloes with the highest \Vmax{}, which is consistent with expectations from galaxy formation models. At low \Vmax{}, the limited resolution of the simulation prevents the full characterization of the low-mass halo population, producing a turnover in the \Vmax{} functions below \Vmax[{\kms[4]}]. In this regime, the suppression of low-\Vmax{} backsplash and \hermeian{} haloes is enhanced because of their repeated interactions with the Local Group primaries. The tidal stripping they experience removes enough material during the pericentric passages that low-\Vmax{} haloes stop being identified by structure finders or are numerically disrupted \citep[see discussion in e.g.][]{springel_aquarius_2008,penarrubia_impact_2010,onions_subhaloes_2012,van_den_bosch_dissecting_2017,newton_total_2018,van_den_bosch_dark_2018,green_tidal_2019,errani_asymptotic_2021,poulton_extracting_2020,green_tidal_2021}, which manifests partly as a low-\Vmax{} truncation of the distributions. This feature moves towards higher values of \Vmax{} with each pericentric passage a halo population experiences, as formerly high-\Vmax{} haloes are repeatedly restructured, stripped of material, and are eventually destroyed by numerical effects.

The upper right-hand panel of \figref{fig:Results:c200_vs_vmax_vmax_vs_m200} shows that the total halo mass (\Mvir{}) functions of the four halo populations are consistent with each other across the mass range at \z[0]. Although resolution effects are particularly severe for low-mass haloes in all of the populations, they also affect more massive haloes in the interacting populations that experience considerable tidal stripping after accretion into a host \citep[see e.g.][]{newton_constraints_2021}. In comparison, the limited resolution affects the \Vmax{} of haloes across a much narrower range of values (upper left-hand panel). This is because the circular velocity curves close to the centres of the haloes remain relatively untouched by repeated tidal stripping episodes that mostly affect the outer parts of the haloes. As we discussed above, the turnover at \Vmax[{\kms[4]}] is caused by the numerical destruction of some low-mass and highly stripped haloes. Those that are left are the remnants of more concentrated, higher mass progenitors at infall that survive repeated interactions with the primary haloes. In a higher resolution simulation many of the destroyed haloes would survive to \z[0] as low-mass DM remnants devoid of baryonic material \citep[see e.g.][]{green_tidal_2021}.

In the lower right-hand panel, we plot the median \Vmax{}--\Mvir{} relation of the surviving haloes. These relations are tightly correlated across several orders of magnitude in mass and indicate the concentrations of the haloes without assuming a particular form for their DM density profiles. The \Vmax{}--\Mvir{} relationship of the regular field haloes is consistent with the \cvir{}--\Mvir{} relation from \citet{ludlow_massconcentrationredshift_2014}. At fixed mass, haloes that interact with a primary halo have higher \Vmax{} compared to the regular field haloes and are therefore more concentrated. Of the haloes that interact with the primaries, those that experience two pericentric passages are approximately twice as concentrated as the regular field haloes, and the \Nperi[1] backsplash haloes are a factor of $1.7$~times more concentrated. The destruction of low-\Vmax{} haloes in the interacting halo populations is enhanced compared with the regular field haloes, which biases our results towards higher \Vmax{} values and higher halo concentrations. This effect is larger at the low-mass end but is almost completely absent at higher masses. As the interacting halo populations have higher \Vmax{} across the entire range in halo mass, we think that this systematic effect is small and does not affect our conclusions.

\subsection{Conduits of matter transfer between primary haloes}
\label{sec:Results:Particle_exchange}
\begin{figure}%
    \centering%
	\includegraphics[width=\columnwidth]{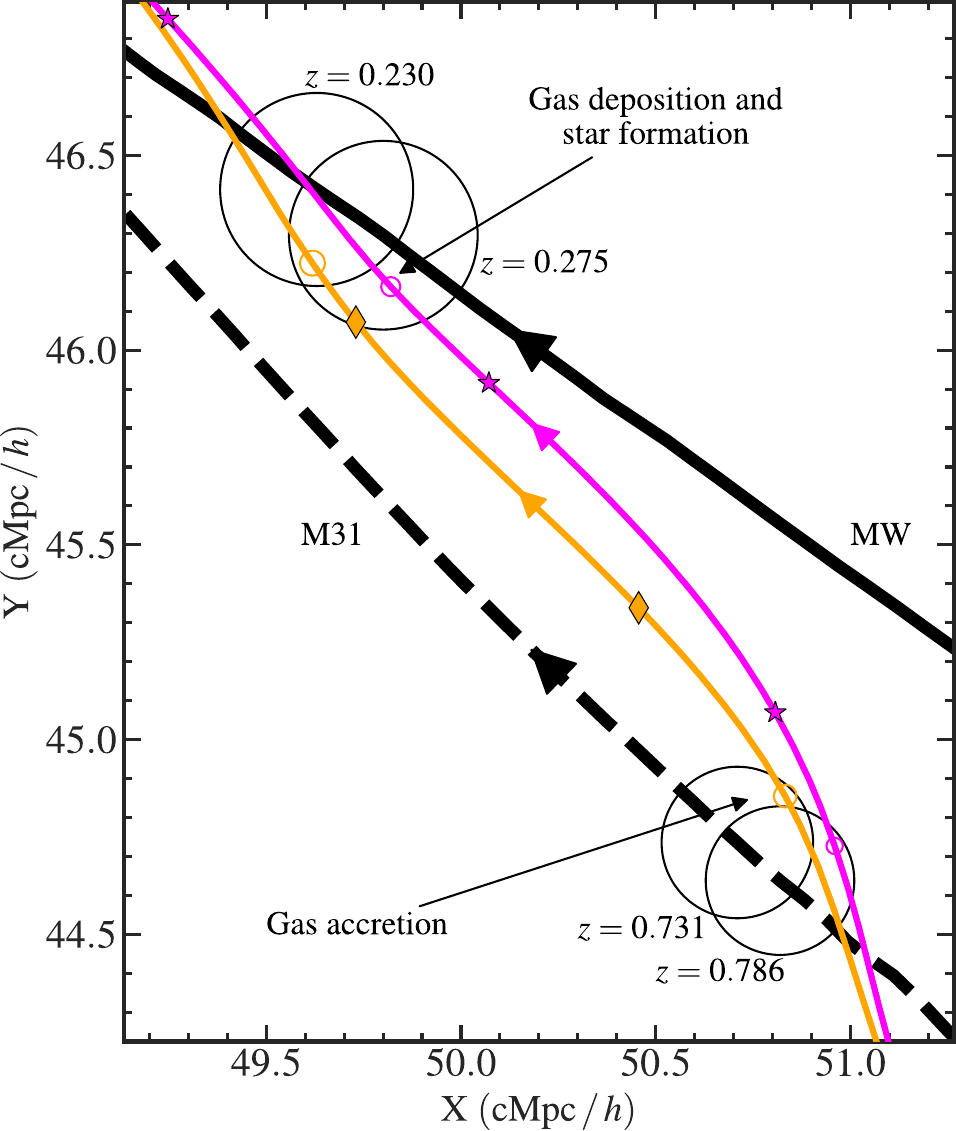}%
	\caption{Illustration of the transfer of gas, and the subsequent formation of stars, from one primary halo to the other facilitated by \hermeian{} galaxies. This shows the trajectories of the MW and M31 analogues (thick solid and dashed lines, respectively) and two \hermeian{} galaxies (solid lines with distinguishing markers as described in \figref{fig:Methods:hermeian_distance}) projected onto the simulation co-moving X--Y plane. Arrow heads indicate the directions of travel of the haloes, and the circles show the \Rvir{} boundaries of the haloes at redshifts, $\z[{\left[0.786, 0.731, 0.275, 0.230\right]}]$. Gas accretion/deposition by the \hermeian{} galaxies and the resulting star formation events in the primary halo are marked on the figure.
	}%
	\label{fig:Results:gas_sf_transfer}%
	\vspace{-10pt}%
\end{figure}%

When the \hermeian{} haloes first encounter a primary halo some of their DM and baryonic material is stripped away. During this interaction they can also accrete matter from the host and carry it with them as they re-enter the field, later depositing it in the second host halo as they pass through. Most of the mass transferred between the primaries in this manner is in the form of DM. In the \code{17_11} simulation, nearly $\Msun[6\times10^8]$ of the MW DM halo at \z[0] was deposited via \hermeian{} transfer from the M31 analogue; similarly, almost $\Msun[5\times10^6]$ of the M31 DM halo originated in the MW analogue and was transferred by the same mechanism.

The rest of the mass exchanged via \hermeian{} haloes is baryonic and in the form of gas and stars. In the \code{17_11} simulation, this is transferred between the two hosts by the two most massive \hermeian{} galaxies. The gas is chemically similar to the composition of the primary it originated from at the early times when it was accreted by the \hermeian{} halo. While being transferred to the other primary the gas and the \hermeian{} galaxy exchange a small amount of material. This could leave an observable chemical signature that distinguishes the \hermeian{} galaxy from other field dwarf galaxies. Furthermore, when the gas is deposited into the second host at late times it is a source of chemical pollution in that halo that carries chemical signatures from both the other primary halo and the \hermeian{} halo that transferred it. The \hermeian{} haloes transfer $\Msun[2.2\times10^7]$ of gas from the M31 analogue to the MW halo but none in the opposite direction. The accretion and deposition events of the two \hermeian{} haloes are separated by \Myr[290] and \Myr[450], respectively, and so are relatively close in time to one another.

Occasionally, during the deposition of gas into the primary haloes its pressure increases substantially, which increases the star formation rate \citep{grand_gas_2019}. In \figref{fig:Results:gas_sf_transfer}, we show the trajectories, projected into the simulation comoving X--Y plane, of the MW and M31 analogues and the two \hermeian{} galaxies that facilitate the star formation events. The stellar winds that are produced enrich the surrounding interstellar medium of the primary halo. These interactions demonstrate that some low-mass stellar populations in the MW or M31 could have been seeded by gas exchanged via a \hermeian{} galaxy. This could explain the origin of some stellar streams and stars with unusual chemical compositions \citep[e.g.][]{xing_evidence_2019,wan_tidal_2020}.

Finally, we also find that \hermeian{} galaxies can accrete stars from the primary haloes and transport them through the Local Group. In this simulation, the stars are stripped from the \hermeian{} galaxy during its interaction with the second primary halo, becoming part of the tidal debris stretching into the field in the wake of the \hermeian{} galaxy. At \z[0], the stars are \kpc[430] from the closest primary halo and \kpc[100] from the \hermeian{} galaxy that transported them. Despite their turbulent history, the stars are not chemically polluted by either of the galaxies that they were in contact with.

We note that these results emerge by tracking the exchange of a very small sample of star and gas particles between haloes, so we cannot make quantitative statements about the size of any chemical pollution signal or the possibility of using it as a tracer of these interactions. Our results are also subject to the limited resolution of the simulation and the assumptions incorporated into the subgrid physics models \citep[see][for a review of galaxy formation models in simulations]{vogelsberger_cosmological_2020}. Higher resolution simulations, an improved understanding of the processes currently encapsulated in the subgrid model, and future developments in the implementation of the hydrodynamic scheme will enable a more thorough exploration of this subject. Therefore, we present this result -- that there exists a population of haloes currently in the field that may have acted as matter conduits between the MW and M31 -- as a `proof of concept' that will benefit from further study in future more targeted work.

\subsection{Prospects for indirect DM searches}
\label{sec:Results:Jfactors}
\begin{figure*}%
    \centering%
	\includegraphics[width=\textwidth]{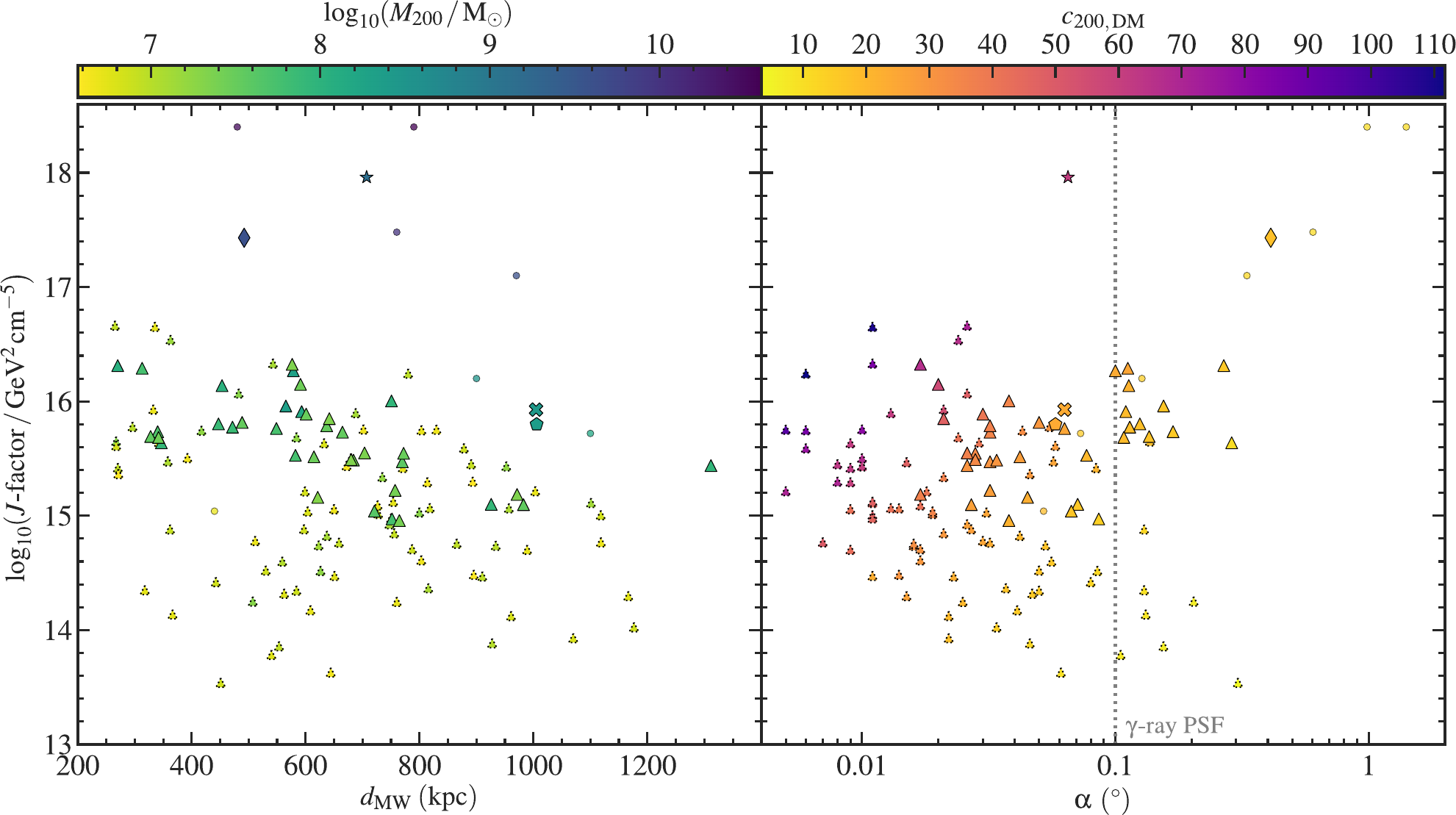}%
	\caption{\textit{Left-hand panel:} Astrophysical \Jfactor{}s of the \hermeian{} haloes as a function of distance from the MW analogue coloured by halo mass, \Mvir{}. \textit{Right-hand panel:} \Jfactor{}s of the \hermeian{} haloes as a function of the angular sizes, $\alpha$, of their photon emission regions coloured by the concentration, \conc{{\rm 200,\, DM}}, of the DM component of the halo. The vertical dotted line indicates the typical angular resolution, or Point Spread Function~(PSF), of current gamma-ray experiments such as \Fermi{} or MAGIC. Objects to the left of this line will appear as point-like sources in observations. In both panels, haloes with masses below that of $100$~DM particles are plotted with a dotted outline. Additionally, the four \hermeian{} galaxies are represented by the same markers as in \figref{fig:Methods:hermeian_distance}. For comparison, we also plot as points the results of halo modelling based on observations of seven Local Group field galaxies \citep{gammaldi_dark_2021}.}%
	\label{fig:Results:J_factors}%
	\vspace{-10pt}%
\end{figure*}%

Highly concentrated DM-dominated objects such as the \hermeian{} haloes can be excellent targets for indirect DM searches that look for DM annihilation products. In DM models where the particle annihilates and produces photons, the expected photon flux from a DM source, $\tfrac{d\Phi_\gamma}{dE}$, can be expressed as the product of two physically motivated terms:
\begin{equation}
    \label{eq:Results:DM_searches:Photon_flux}
    \frac{d\Phi_\gamma}{dE} = \frac{d\Phi^{pp}_\gamma}{dE} \times J\!\left(\Delta\Omega\right)\,.
\end{equation}
The first term, $\tfrac{d\Phi^{pp}_\gamma}{dE}$, contains information about the DM particle properties and its interactions (e.g. its mass, annihilation cross-section and number of photons per annihilation) and, for Majorana DM particles, can be expressed as
\begin{equation}
    \label{eq:Results:DM_searches:Photon_PP_flux}
    \frac{d\Phi^{pp}_\gamma}{dE} = \frac{\langle\sigma_A v\rangle}{8 \pi m^2_{\rm DM}} \frac{dN_\gamma}{dE}\,,
\end{equation}
where $m_{\rm DM}$ is the mass of the DM particle, $\langle\sigma_A v\rangle$ is the thermally averaged DM annihilation cross-section, and $\tfrac{dN_\gamma}{dE}$ describes the energy distribution of the photons produced by the interactions. The second term in \eqnref{eq:Results:DM_searches:Photon_flux}, called the \Jfactor{}, $J\!\left(\Delta\Omega\right)$, encodes the astrophysical properties of the DM in the target system. If the DM particle interaction cross-section does not depend on the particle velocity, then the \Jfactor{} is independent of the DM model,\footnote{In detail, there is always a connection between both terms in \eqnref{eq:Results:DM_searches:Photon_flux}: the mass of the DM particle imposes a cut-off on the matter power spectrum that sets a minimum halo mass, $M_{\rm min}$. For DM particles with $m_{\rm DM} = \GeV[\bigO{100}]$, the minimum halo mass is approximately ${M_{\rm min}=\Msun[10^{-6}]}$~\citep{profumo_what_2006,bringmann_particle_2009}. The value of this quantity has important implications for the total annihilation signal \citep{sanchez-conde_flattening_2014,moline_characterization_2017}.} making it useful to assess the suitability of \hermeian{} haloes as targets for indirect DM searches. The \Jfactor{} is calculated by
\begin{equation}
    \label{eq:Results:DM_searches:J_factor}
    J\!\left(\Delta\Omega\right) = \int_{\Delta\Omega} \int_{l.o.s.} \rho^2_{\rm DM}\!\left(\textbf{r}\!\left(l,\, \Omega\right)\right)\, dl d\Omega\,,
\end{equation}
where $\rho_{\rm DM}$ is the radial-dependent DM density profile of the DM target, \textit{l} is the distance along the line of sight, and $\Delta\Omega$ is the solid angle of the observation \citep{evans_travel_2004}. Assuming a NFW DM density profile, the {\it total} \Jfactor{} integrated up to \Rvir{} of the target, $J_T$, can be written as a function of both $M_{200}$ and $c_{200}$ as
\begin{equation}
\label{eq:Results:DM_searches:J_factor_NFW}
\begin{aligned}
    J_{T} &= \frac{1}{4\pi d^2}\int_{V} \rho^2_{\rm DM}\!(\textbf{r})~dV \\
    &= \frac{1}{d^2} \frac{M_{200} \,c_{200}^3}{\left[f(c_{200})\right]^{2}} \frac{200 \, \rho_{\rm crit}}{9} \, \left(1-\frac{1}{(1+c_{200})^3} \right)\,,
\end{aligned}
\end{equation}
where \textit{d} is the distance from the Earth to the centre of the target system and $f(x)= \mbox{ln}(1+x)-x/(1+x)$. To quantify the strength of the annihilation signal from the \hermeian{} haloes, we calculate their total \Jfactor{}s by modelling their DM components as NFW profiles. We exclude the baryons because they do not contribute to the annihilation signal, so the halo concentrations we obtain in this section differ from those in \figref{fig:Results:c200_vs_vmax_vmax_vs_m200} that are calculated by following the prescription in \citet{prada_halo_2012}, which assumes a NFW profile form. In \figref{fig:Results:J_factors}, we plot the \Jfactor{}s as a function of the distance from the MW analogue (left-hand panel) and of the angular size of the photon emission region (right-hand panel), which we define as the angle subtended by twice the scale radius of the target. When assuming a NFW density profile, approximately \percent{90} of the annihilation flux originates within this region and it therefore represents a good proxy of the angular size that would be observed by gamma-ray telescopes. Recently, \citet{wang_universal_2020} showed that an Einasto DM density profile with shape parameter, $\alpha = 0.16$, fits the profiles of DM haloes more accurately than a NFW profile for halo masses spanning $20$ orders of magnitude. We expect that repeating the above calculations assuming an Einasto profile would produce very similar results.

When calculating the \Jfactor{} we do not include any enhancement in the signal due to the subhalo population inside the \hermeian{} halo \citep[the so-called {\it subhalo boost}, e.g.][]{kamionkowski_galactic_2010, sanchez-conde_flattening_2014,ando_halo_2019}. Having been subhaloes themselves in the past, \hermeian{} haloes experienced strong tidal interactions with both the MW and M31. This repeated tidal stripping not only increased their concentrations but also will have removed most of their substructure, especially in the outermost regions. As a result, the subhalo boost is expected to be of the order of only a few percent for these objects~\citep{moline_characterization_2017}.

The average total \Jfactor{} of the dark \hermeian{} population resolved with at least 100~DM particles is ${\log_{10}\left(\tilde{J}_T\, /\, \GeV{}^2\unit{cm}^{-5}\right)=}$$15.69^{+0.35}_{-0.51}$ and the four \hermeian{} galaxies have ${\log_{10}\left(J_T\, /\, \GeV{}^2\unit{cm}^{-5}\right)=}$$\left[15.80,\, 15.93,\, 17.43,\, 17.96\right]$, respectively. The \Jfactor{} is affected most strongly by the halo mass and is weakly sensitive to the distance of the \hermeian{} haloes from the primary halo. Consequently, $8{-}13$ of the dark haloes resolved with at least $100$~DM particles have larger \Jfactor{}s than the two low-mass \hermeian{} galaxies. The maximum \Jfactor{} among the \hermeian{}s reaches a value of ${\log_{10}\left(J_T/\GeV{}^2\unit{cm}^{-5}\right)=}17.96$. This is comparable to the \Jfactor{}s of other more traditional targets for gamma-ray DM searches, such as dwarf satellite galaxies, dark satellites, or galaxy clusters \citep{charles_sensitivity_2016}. However, this is significantly below the \Jfactor{}s of the most promising nearby MW dwarf galaxies~\citep{albert_searching_2017} that are usually considered to be among the best targets for indirect DM detection, though see the recent discussions in \citet{facchinetti_statistics_2020} and \citet{grand_baryonic_2021} about the likelihood of detecting an unambiguous signal in the smooth halo of the MW. Detecting signals from distant field galaxies is more challenging and to date there have been few analyses of such objects. Recently, \citet{gammaldi_dark_2021} modelled the DM distributions of seven dwarf irregular galaxies in the field of the Local Group using observations from \Fermi{}. The \Jfactor{}s they calculate for these galaxies are consistent with those of \hermeian{} haloes with similar masses; however, the field galaxies are less concentrated than most of the \hermeian{} haloes and so their photon emission regions are more spatially extended. We expect that the more concentrated \hermeian{} haloes will provide stronger signals compared to regular field haloes of a similar mass that are at a similar distance. We compare the annihilation signals from the field halo populations in \appref{sec:Appendix:17_11_Jfactors}.

In addition to the \Jfactor{}, the detectability of DM haloes in gamma-ray searches also depends on the angular size, $\alpha$, of the photon emission region. This is because the analysis of gamma-ray data from extended sources is significantly more complex than that for point-like sources \citep[see the discussion in e.g.][]{ackermann_search_2015,acciari_constraining_2018}. The angular size depends on both the distance to the source and the spatial extent of its photon emission, which scales inversely with halo concentration. In the right-hand panel of \figref{fig:Results:J_factors}, most \hermeian{} haloes have angular sizes ${\alpha < 0.1^\circ}$ that are smaller than the typical angular resolution of current gamma-ray experiments such as \Fermi{} \citep{atwood_large_2009} and the Major Atmospheric Gamma Imaging Cherenkov~(MAGIC) telescope \citep{aleksic_major_2016}. Consequently, these objects will appear as point-like sources in gamma-ray telescopes and are promising candidates to detect a DM signal. The most extended \hermeian{} haloes have angular sizes ${\alpha < 0.4^\circ}$, which is still small enough to have only a minor effect on analyses to detect DM signals. Indeed, these angular sizes are smaller than or comparable to those of dwarf galaxies and are more than $10$ times smaller than the typical angular size subtended by the most promising galaxy clusters for DM detection \citep{sanchez-conde_dark_2011}. We note though that the brightest \hermeian{}s would appear as extended sources for the future Cherenkov Telescope Array~(CTA), which will have an angular resolution better than $0.05^\circ$ at energies above \TeV[1]~\citep{acharya_introducing_2013,the_cta_consortium_science_2019}. We also note that, in detail, the detectability of objects in a given instrument depends on fore-/background modelling assumptions \citep[see e.g.][]{sanchez-conde_dark_2011,bonnivard_dark_2015}, the characteristics of the instrument, and the photon flux from the source, which depends on the particle properties of the DM.

The size of the \hermeian{} population could also prove to be advantageous when conducting gamma-ray data analyses. If it is as large as presented here, the sensitivity of such analyses to a DM signal could be enhanced significantly by carrying out a combined likelihood analysis on the full sample of \hermeian{} haloes, as is done for dwarf galaxies~\citep{ackermann_searching_2015}. In practice, such combined analyses would be dominated by the objects with the highest \Jfactor{}s. However, as we show in \figref{fig:Results:J_factors}, most \hermeian{} haloes have similarly high \Jfactor{}s and therefore many haloes would contribute to increase the statistical power in such a combined DM search.

\section{Discussion and Conclusions}
\label{sec:Disc_Conc}
We have discovered and described a new class of cosmic objects that passed within the virial radii of {\it both} Local Group hosts, the MW and M31, and became field haloes at \z[0]. We call these `\hermeian{}' haloes, named after Hermes from Ancient Greek mythology: the messenger of the gods, who presided over the crossing of thresholds and boundaries. These \hermeian{} haloes are similar to `backsplash' haloes that experience more than one pericentric passage with the same host; however, \hermeian{} haloes pass through two different and massive haloes at two different epochs in their formation. Most \hermeian{} haloes are devoid of baryonic material at \z[0] as the gas and stellar content is removed during their pericentric passages. The most massive \hermeian{} haloes can contain galaxies that survive close interactions with the primary hosts and retain their gaseous component. These also facilitate the exchange of DM, gas, and stars between the primary haloes and can also trigger star formation during their interactions with the two hosts. Such \hermeian{} galaxies in the Local Group would appear as gas-poor field galaxies aligned approximately along the line connecting the MW and M31, and would have distinct chemical signatures of their interactions with the MW and M31 at early times. \hermeian{} haloes are therefore important participants in the evolution of the MW and M31 that affect observational proxies of the star formation histories of the main galaxies. These intergalactic messengers are potentially a source of external chemical pollution within a galactic DM halo that we will investigate in future studies.

The set of trajectories that \hermeian{} haloes take through the Local Group is limited. This restricts their spatial distribution at \z[0] and favours a highly anisotropic configuration along the line connecting the MW with M31 (see \figref{fig:Results:hermeian_midpoint_angle}). We find \hermeian{} haloes \kpc[400] to \Mpc[1.6] from the midpoint of this connecting line, and none between the primary hosts. That a halo at such large distances could have passed through both primary haloes without being accreted or dissolved by tidal stripping is particularly remarkable and speaks to the resilience of some haloes. When projected on the sky, the \hermeian{} haloes appear to cluster close to the projected positions of the MW and M31 (see \figref{fig:Results:hermeian_aitoff}). The tightly constrained spatial distribution will assist searches for observational signatures of the dark haloes, and also can be applied as a criterion to select candidate \hermeian{} galaxies for additional follow-up. Further discoveries could be made in upcoming wide-field imaging surveys such as the Legacy Survey of Space and Time at the Vera C. Rubin Observatory \citep{ivezic_lsst_2019}.

The properties of \hermeian{} haloes are affected by tidal stripping and shock heating during their interactions with the Local Group primaries. These effects are strongest during pericentric passage and affect the \Vmax{} distribution and concentrations of the \hermeian{} population (see \figref{fig:Results:c200_vs_vmax_vmax_vs_m200}). Compared with the regular field halo population (i.e. excluding the \hermeian{} and backsplash haloes), the \hermeian{} haloes sample a narrower range in \Vmax{}. The truncation of the upper end of the distribution to \Vmax[{\kms[35]}] is a consequence of tidal stripping that removes mass efficiently from the outer layers of the haloes. This also affects the low-\Vmax{} end, where haloes are also more susceptible to artificial disruption by numerical effects in the simulation that suppress the abundance of field haloes below \Vmax[{\kms[4]}]. This turnover is more severe in the backsplash and \hermeian{} halo populations as tidal interactions dissolve some low-\Vmax{} haloes during each pericentric passage. The same mechanisms also enhance the concentrations of the haloes while they travel through the Local Group hosts, changing the shape of the \hermeian{} concentration distribution function and making them almost twice as concentrated, on average, as the regular field haloes. We find a similar result for the population of backsplash haloes that passed through a primary halo twice: on average, they are \percent{14} more concentrated than the \hermeian{} population and more than twice as concentrated as the regular field haloes. However, their spatial distribution in the Local Group at \z[0] is less anisotropic than the \hermeian{} haloes, which makes searching for `\Nperi[2]' backsplash haloes in observations more challenging.

In the \code{17_11} \Hestia{} simulation most of the \hermeian{} haloes are created as part of a large group that falls into the M31 analogue at early times. This group travels through the Local Group from the direction of M31 towards the MW so that, at \z[0], the angular distribution of \hermeian{} haloes in the direction of the MW analogue (\thetamid[{180^\circ}]) is enhanced. The remaining non-\hermeian{} haloes in the infalling group have three possible futures:
\begin{enumerate*}
    \item they avoid interacting with either host and remain as field haloes;
    \item they are captured by one of the host haloes and become a satellite, or;
    \item they experience at least one pericentric passage with a host and become backsplash haloes.
\end{enumerate*}
The backsplash haloes that fell in are spread throughout the Local Group at \z[0]; however, most cluster close to the M31 primary, which enhances the angular distribution of backsplash haloes towards \thetamid[{0^\circ}] (see \figref{fig:Results:hermeian_midpoint_angle}). The asymmetric enhancements of the angular distributions of the backsplash and \hermeian{} populations towards opposing poles of the MW--M31 line are a characteristic signature of the group infall. Evidence of similar major encounters in the MW or M31, for example from starburst events or other chemical signatures, and the dynamical modelling of their nearby dwarf galaxies will help to establish the abundance and likely locations of \hermeian{} haloes in the Local Group \citep[e.g.][]{teyssier_identifying_2012,buck_nihao_2019,mcconnachie_solo_2021}. The members of the NGC~$3109$ association are interesting targets for further study because initial calculations suggest these galaxies could have experienced interactions with the MW, indicating that they are remnants of a similar group infall event \citep{shaya_formation_2013,pawlowski_perseus_2014}. However, better proper motion measurements and additional dynamical modelling are needed to establish whether the association has also interacted with M31.

\hermeian{} galaxies can facilitate the exchange of DM and baryonic material between the two primary haloes. This seeds the MW and M31 with gas that is chemically similar to the primary halo it originated from but distinct from the \hermeian{} galaxy in which it travels and the halo into which it is deposited. This could imprint a chemical signature in the composition of the gas in the \hermeian{} galaxy that distinguishes it from field dwarf galaxies that have not interacted with the MW or M31. As the \hermeian{} galaxies pass through the primary haloes, the star formation rate of the gas they deposit increases, which incorporates the polluted gas into new stars that have unusual chemical compositions when compared with the surrounding interstellar medium and the rest of the stellar population of the host (see \figref{fig:Results:gas_sf_transfer}). \hermeian{} galaxies can also transport stars from one primary halo to another. In the \code{17_11} simulation, these stars are stripped from the \hermeian{} galaxy and form part of the tidal debris left in its wake as it passes through the second primary halo and back into the field. At \z[0] the stars are \kpc[430] from this primary halo and \kpc[100] from the \hermeian{} galaxy.

The \hermeian{} haloes are promising targets to detect signals of DM annihilation in gamma-ray experiments compared to field galaxies with a similar mass. We characterize their detectability by calculating their astrophysical \Jfactor{}s and determine the angular sizes of their photon emission regions (see \figref{fig:Results:J_factors}). The \Jfactor{}s of the four \hermeian{} galaxies are comparable with those from dwarf galaxies but are much lower than the nearest MW satellite galaxies showing most promise for the detection of a DM signal. However, most of the \hermeian{} haloes would be point-like sources in current gamma-ray telescopes, which improves their detectability significantly. If the real \hermeian{} population is as large as the simulated one, a combined likelihood analysis of gamma-ray data would increase the sensitivity to DM signals from these objects. In principle, the sensitivity could be increased further by carrying out a combined likelihood analysis of the \Nperi[2] backsplash haloes, which are similarly highly concentrated and typically are closer to the host haloes.
This would require a more intensive, whole-sky search to identify the full population of \Nperi[2] backsplash galaxies because they are closer and are distributed approximately isotropically with respect to the Milky Way.
In contrast, most of the population of \hermeian{} haloes are located in two lobes aligned along the Milky Way--M31 line, so the \hermeian{} galaxies could be found more efficiently using smaller targeted surveys.

We find \hermeian{} haloes in all three high-resolution simulated Local Groups in the \Hestia{} suite. However, over \percent{90} of all the \hermeian{} haloes in these simulations, as well as all the \hermeian{} haloes that retain baryons at \z[0], are in just one \Hestia{} simulation (\code{17_11}). This Local Group has the most massive primary pair that is also \percent{20} closer together at \z[0] than in the other simulations. We note that the primaries in this simulation have masses consistent with observational constraints and a separation at \z[0] of \kpc[675], which is slightly less than the observed value of \kpc[{\sim}780] \citep{mcconnachie_distances_2005,conn_bayesian_2012,riess_cepheid_2012}. This suggests that the masses and spatial evolution of the Local Group primaries influence the creation of \hermeian{} galaxies. This is consistent with the results of \citet{knebe_renegade_2011}, who found that the closer proximity of the hosts during their evolution enhances the rate of exchange of low-mass satellites. We also show that the abundance of \hermeian{} haloes can be affected by group infall, which depends on the location and orientation of the primary haloes with respect to the Large Scale Structure. The size of the \hermeian{} population varies between the three high-resolution simulations of the \Hestia{} suite, although each simulation is consistent with observations of the large-scale cosmography and the geometric and dynamical properties of the MW and M31. Therefore, we expect that the Local Group will contain \hermeian{} haloes that have influenced the assembly of the MW and M31, although they are likely to be dark structures devoid of baryons at \z[0]. If they are present, \hermeian{} galaxies will be very rare but could be distinguished from other field galaxies by chemical signatures that allude to their past interactions with the MW and M31. These gas-poor galaxies, like the dark \hermeian{} haloes, would lie preferentially along the line connecting the MW and M31. A quantitative statement about the Local Group properties that strongly affect the size of the \hermeian{} population would require the analysis of a large suite of more highly resolved constrained Local Group simulations, which we defer to future work.

\hermeian{} haloes are a universal feature of paired- and multi-halo systems, like the Local Group and massive clusters. Although rare, their high concentrations make them promising targets to detect DM annihilation signals. The highly anisotropic spatial distribution of the \hermeian{} haloes will assist in the search for these objects in observational campaigns, and analyses of DM annihilation signals could be enhanced further by carrying out a combined likelihood analysis of the separate sources. The mass function of the \hermeian{} population is similar to that of the regular field haloes, suggesting that there could be many \hermeian{} haloes that are not resolved in current simulations. The low-mass haloes that are resolved are more susceptible to artificial disruption from numerical effects, a process that is exacerbated during interactions with massive hosts. Therefore, higher resolution hydrodynamic simulations of constrained Local Group systems are needed to estimate the abundance of \hermeian{} haloes and galaxies in the Local Group. They will also help to understand better how \hermeian{} haloes affect the evolution of the baryonic component of the MW and M31 haloes and their satellites, and will also shed light on the potential use of the \hermeian{} haloes for indirect DM searches. Some galaxies, such as those in the NGC~$3109$ association, are promising \hermeian{} candidates.

\section*{Acknowledgements}

We thank the anonymous referee for detailed comments that improved the quality of the manuscript.
We also thank Rob Grand, Christoph Pfrommer, Marcel Pawlowski, and Stefan Gottl\"{o}ber for useful discussions on the draft manuscript.
The authors gratefully acknowledge the Gauss Centre for Supercomputing e.V. (\url{www.gauss-centre.eu}) for supporting the \Hestia{} project by providing computing time on the GCS Supercomputer SuperMUC-NG at Leibniz Supercomputing Centre (\url{www.lrz.de}). We also acknowledge use of the data storage system EREBOS at AIP.
ON thanks Tom Rose for directing attention to Hermes from Ancient Greek mythology and his grammatical comments on the manuscript. ON also thanks Tom Rose, Thomas Callingham, and Calvin Sykes for their hospitality during the completion of this work. ON and NIL acknowledge financial support from the Project IDEXLYON at the University of Lyon under the Investments for the Future Program (ANR-16-IDEX-0005) and supplementary financial support from La R\'{e}gion Auvergne-Rh\^{o}ne-Alpes. ON acknowledges additional financial support from the Royal Society.
AK is supported by the Ministerio de Ciencia, Innovaci\'{o}n y Universidades (MICIU/FEDER) under research grant PGC2018-094975-C21 and further thanks Low for \textit{I Could Live in Hope}.
MASC acknowledges the support of the {\it Atracci\'on de Talento Investigador} contract no. 2020-5A/TIC-19725 granted by the Comunidad de Madrid in Spain. MASC was additionally supported by the Spanish Agencia Estatal de Investigación through the grant PGC2018-095161-B-I00 and the IFT Centro de Excelencia Severo Ochoa, SEV-2016-0597.
ET acknowledges support by ETAg grant PRG1006 and by the EU through the ERDF CoE grant TK133.

\textit{Software}: This research made use of \astropy{} \citep{the_astropy_collaboration_astropy_2013,the_astropy_collaboration_astropy_2018}, \matplotlib{} \citep{hunter_matplotlib_2007}, \numpy{} \citep{walt_numpy_2011,harris_array_2020}, \python{} \citep{van_rossum_python_2009}, \scipy{} \citep{jones_scipy_2011,virtanen_scipy_2020}, and the NASA Astrophysics Data System. We thank their developers for maintaining them and making them freely available.

\section*{Data Availability}

The data used in this work were extracted from the \Hestia{} simulation suite. A data repository and scripts to produce the figures in this manuscript are available on GitHub
\footnote{Supplementary materials: \url{https://github.com/Musical-Neutron/hermeian_paper_plots/}}
and archived in Zenodo \citep{newton_hermeian_2022-1}.
Requests for access to the \Hestia{} simulation data should be directed to a CLUES Collaboration PI.



\bibliographystyle{mnras}
\bibliography{HESTIA_hermeian} 

\begin{thebibliography}{}
\makeatletter
\relax
\def\mn@urlcharsother{\let\do\@makeother \do\$\do\&\do\#\do\^\do\_\do\%\do\~}
\def\mn@doi{\begingroup\mn@urlcharsother \@ifnextchar [ {\mn@doi@}
  {\mn@doi@[]}}
\def\mn@doi@[#1]#2{\def\@tempa{#1}\ifx\@tempa\@empty \href
  {http://dx.doi.org/#2} {doi:#2}\else \href {http://dx.doi.org/#2} {#1}\fi
  \endgroup}
\def\mn@eprint#1#2{\mn@eprint@#1:#2::\@nil}
\def\mn@eprint@arXiv#1{\href {http://arxiv.org/abs/#1} {{\tt arXiv:#1}}}
\def\mn@eprint@dblp#1{\href {http://dblp.uni-trier.de/rec/bibtex/#1.xml}
  {dblp:#1}}
\def\mn@eprint@#1:#2:#3:#4\@nil{\def\@tempa {#1}\def\@tempb {#2}\def\@tempc
  {#3}\ifx \@tempc \@empty \let \@tempc \@tempb \let \@tempb \@tempa \fi \ifx
  \@tempb \@empty \def\@tempb {arXiv}\fi \@ifundefined
  {mn@eprint@\@tempb}{\@tempb:\@tempc}{\expandafter \expandafter \csname
  mn@eprint@\@tempb\endcsname \expandafter{\@tempc}}}

\bibitem[\protect\citeauthoryear{Acciari et~al.,}{Acciari
  et~al.}{2018}]{acciari_constraining_2018}
Acciari V.~A.,  et~al., 2018, \mn@doi [Physics of the Dark Universe]
  {10.1016/j.dark.2018.08.002}, 22, 38

\bibitem[\protect\citeauthoryear{Acharya et~al.,}{Acharya
  et~al.}{2013}]{acharya_introducing_2013}
Acharya B.~S.,  et~al., 2013, \mn@doi [Astroparticle Physics]
  {10.1016/j.astropartphys.2013.01.007}, 43, 3

\bibitem[\protect\citeauthoryear{Ackermann et~al.,}{Ackermann
  et~al.}{2015a}]{ackermann_searching_2015}
Ackermann M.,  et~al., 2015a, \mn@doi [Phys. Rev.]
  {10.1103/PhysRevLett.115.231301}, 115, L231301

\bibitem[\protect\citeauthoryear{Ackermann et~al.,}{Ackermann
  et~al.}{2015b}]{ackermann_search_2015}
Ackermann M.,  et~al., 2015b, \mn@doi [ApJ] {10.1088/0004-637X/812/2/159}, 812,
  159

\bibitem[\protect\citeauthoryear{Ackermann et~al.,}{Ackermann
  et~al.}{2017a}]{ackermann_observations_2017}
Ackermann M.,  et~al., 2017a, \mn@doi [ApJ] {10.3847/1538-4357/aa5c3d}, 836,
  208

\bibitem[\protect\citeauthoryear{Ackermann et~al.,}{Ackermann
  et~al.}{2017b}]{ackermann_fermi_2017}
Ackermann M.,  et~al., 2017b, \mn@doi [ApJ] {10.3847/1538-4357/aa6cab}, 840, 43

\bibitem[\protect\citeauthoryear{Albert et~al.,}{Albert
  et~al.}{2017}]{albert_searching_2017}
Albert A.,  et~al., 2017, \mn@doi [ApJ] {10.3847/1538-4357/834/2/110}, 834, 110

\bibitem[\protect\citeauthoryear{Aleksi{\'c} et~al.,}{Aleksi{\'c}
  et~al.}{2016}]{aleksic_major_2016}
Aleksi{\'c} J.,  et~al., 2016, \mn@doi [Astroparticle Physics]
  {10.1016/j.astropartphys.2015.02.005}, 72, 76

\bibitem[\protect\citeauthoryear{Ando, Ishiyama  \& Hiroshima}{Ando
  et~al.}{2019}]{ando_halo_2019}
Ando S.,  Ishiyama T.,   Hiroshima N.,  2019, \mn@doi [Galaxies]
  {10.3390/galaxies7030068}, 7, 68

\bibitem[\protect\citeauthoryear{Atwood et~al.,}{Atwood
  et~al.}{2009}]{atwood_large_2009}
Atwood W.~B.,  et~al., 2009, \mn@doi [ApJ] {10.1088/0004-637X/697/2/1071}, 697,
  1071

\bibitem[\protect\citeauthoryear{Bakels, Ludlow  \& Power}{Bakels
  et~al.}{2021}]{bakels_pre-processing_2021}
Bakels L.,  Ludlow A.~D.,   Power C.,  2021, \mn@doi [MNRAS]
  {10.1093/mnras/staa3979}, 501, 5948

\bibitem[\protect\citeauthoryear{Bartels, Krishnamurthy  \& Weniger}{Bartels
  et~al.}{2016}]{bartels_strong_2016}
Bartels R.,  Krishnamurthy S.,   Weniger C.,  2016, \mn@doi [Phys. Rev.]
  {10.1103/PhysRevLett.116.051102}, 116, L051102

\bibitem[\protect\citeauthoryear{{Ben{\'i}tez-Llambay}, Navarro, Abadi,
  Gottl{\"o}ber, Yepes, Hoffman  \& Steinmetz}{{Ben{\'i}tez-Llambay}
  et~al.}{2013}]{benitez-llambay_dwarf_2013}
{Ben{\'i}tez-Llambay} A.,  Navarro J.~F.,  Abadi M.~G.,  Gottl{\"o}ber S.,
  Yepes G.,  Hoffman Y.,   Steinmetz M.,  2013, \mn@doi [ApJ]
  {10.1088/2041-8205/763/2/L41}, 763, L41

\bibitem[\protect\citeauthoryear{Besla, Kallivayalil, Hernquist, Robertson,
  Cox, van~der Marel  \& Alcock}{Besla et~al.}{2007}]{besla_are_2007}
Besla G.,  Kallivayalil N.,  Hernquist L.,  Robertson B.,  Cox T.~J.,  van~der
  Marel R.~P.,   Alcock C.,  2007, \mn@doi [ApJ] {10.1086/521385}, 668, 949

\bibitem[\protect\citeauthoryear{Bla{\~n}a, Burkert, Fellhauer, Schartmann  \&
  Alig}{Bla{\~n}a et~al.}{2020}]{blana_dwarfs_2020}
Bla{\~n}a M.,  Burkert A.,  Fellhauer M.,  Schartmann M.,   Alig C.,  2020,
  \mn@doi [MNRAS] {10.1093/mnras/staa2153}, 497, 3601

\bibitem[\protect\citeauthoryear{Bonnivard et~al.,}{Bonnivard
  et~al.}{2015}]{bonnivard_dark_2015}
Bonnivard V.,  et~al., 2015, \mn@doi [MNRAS] {10.1093/mnras/stv1601}, 453, 849

\bibitem[\protect\citeauthoryear{Bringmann}{Bringmann}{2009}]{bringmann_particle_2009}
Bringmann T.,  2009, \mn@doi [New J. Phys.] {10.1088/1367-2630/11/10/105027},
  11, 105027

\bibitem[\protect\citeauthoryear{Buck, Macci{\`o}, Dutton, Obreja  \&
  Frings}{Buck et~al.}{2019}]{buck_nihao_2019}
Buck T.,  Macci{\`o} A.~V.,  Dutton A.~A.,  Obreja A.,   Frings J.,  2019,
  \mn@doi [MNRAS] {10.1093/mnras/sty2913}, 483, 1314

\bibitem[\protect\citeauthoryear{Bullock \& {Boylan-Kolchin}}{Bullock \&
  {Boylan-Kolchin}}{2017}]{bullock_small-scale_2017}
Bullock J.~S.,  {Boylan-Kolchin} M.,  2017, \mn@doi [ARA\&A]
  {10.1146/annurev-astro-091916-055313}, 55, 343

\bibitem[\protect\citeauthoryear{Calore, Cholis, McCabe  \& Weniger}{Calore
  et~al.}{2015}]{calore_tale_2015}
Calore F.,  Cholis I.,  McCabe C.,   Weniger C.,  2015, \mn@doi [Phys. Rev. D]
  {10.1103/PhysRevD.91.063003}, 91, 063003

\bibitem[\protect\citeauthoryear{Carlesi et~al.,}{Carlesi
  et~al.}{2016}]{carlesi_constrained_2016}
Carlesi E.,  et~al., 2016, \mn@doi [MNRAS] {10.1093/mnras/stw357}, 458, 900

\bibitem[\protect\citeauthoryear{Carlson \& Profumo}{Carlson \&
  Profumo}{2014}]{carlson_cosmic_2014}
Carlson E.,  Profumo S.,  2014, \mn@doi [Phys. Rev. D]
  {10.1103/PhysRevD.90.023015}, 90, 023015

\bibitem[\protect\citeauthoryear{Charles et~al.,}{Charles
  et~al.}{2016}]{charles_sensitivity_2016}
Charles E.,  et~al., 2016, \mn@doi [Physics Reports]
  {10.1016/j.physrep.2016.05.001}, 636, 1

\bibitem[\protect\citeauthoryear{Conn et~al.,}{Conn
  et~al.}{2012}]{conn_bayesian_2012}
Conn A.~R.,  et~al., 2012, \mn@doi [ApJ] {10.1088/0004-637X/758/1/11}, 758, 11

\bibitem[\protect\citeauthoryear{Daylan, Finkbeiner, Hooper, Linden, Portillo,
  Rodd  \& Slatyer}{Daylan et~al.}{2016}]{daylan_characterization_2016}
Daylan T.,  Finkbeiner D.~P.,  Hooper D.,  Linden T.,  Portillo S. K.~N.,  Rodd
  N.~L.,   Slatyer T.~R.,  2016, \mn@doi [Physics of the Dark Universe]
  {10.1016/j.dark.2015.12.005}, 12, 1

\bibitem[\protect\citeauthoryear{Di~Mauro \& Winkler}{Di~Mauro \&
  Winkler}{2021}]{di_mauro_multimessenger_2021}
Di~Mauro M.,  Winkler M.~W.,  2021, \mn@doi [Phys. Rev. D]
  {10.1103/PhysRevD.103.123005}, 103, 123005

\bibitem[\protect\citeauthoryear{Diemer}{Diemer}{2021}]{diemer_flybys_2021}
Diemer B.,  2021, \mn@doi [ApJ] {10.3847/1538-4357/abd947}, 909, 112

\bibitem[\protect\citeauthoryear{Doumler, Hoffman, Courtois  \&
  Gottl{\"o}ber}{Doumler et~al.}{2013}]{doumler_reconstructing_2013}
Doumler T.,  Hoffman Y.,  Courtois H.,   Gottl{\"o}ber S.,  2013, \mn@doi
  [MNRAS] {10.1093/mnras/sts613}, 430, 888

\bibitem[\protect\citeauthoryear{Enzi et~al.,}{Enzi
  et~al.}{2021}]{enzi_joint_2021}
Enzi W.,  et~al., 2021, \mn@doi [MNRAS] {10.1093/mnras/stab1960}, 506, 5848

\bibitem[\protect\citeauthoryear{Errani \& Navarro}{Errani \&
  Navarro}{2021}]{errani_asymptotic_2021}
Errani R.,  Navarro J.~F.,  2021, \mn@doi [MNRAS] {10.1093/mnras/stab1215},
  505, 18

\bibitem[\protect\citeauthoryear{Evans, Ferrer  \& Sarkar}{Evans
  et~al.}{2004}]{evans_travel_2004}
Evans N.~W.,  Ferrer F.,   Sarkar S.,  2004, \mn@doi [Phys. Rev. D]
  {10.1103/PhysRevD.69.123501}, 69, 123501

\bibitem[\protect\citeauthoryear{Facchinetti, Lavalle  \& Stref}{Facchinetti
  et~al.}{2020}]{facchinetti_statistics_2020}
Facchinetti G.,  Lavalle J.,   Stref M.,  2020, preprint (\mn@eprint {arXiv}
  {2007.10392})

\bibitem[\protect\citeauthoryear{Fattahi et~al.,}{Fattahi
  et~al.}{2016}]{fattahi_apostle_2016}
Fattahi A.,  et~al., 2016, \mn@doi [MNRAS] {10.1093/mnras/stv2970}, 457, 844

\bibitem[\protect\citeauthoryear{Gammaldi, {P{\'e}rez-Romero},
  {Coronado-Bl{\'a}zquez}, Di~Mauro, Karukes, {S{\'a}nchez-Conde}  \&
  Salucci}{Gammaldi et~al.}{2021}]{gammaldi_dark_2021}
Gammaldi V.,  {P{\'e}rez-Romero} J.,  {Coronado-Bl{\'a}zquez} J.,  Di~Mauro M.,
   Karukes E.~V.,  {S{\'a}nchez-Conde} M.~A.,   Salucci P.,  2021, \mn@doi
  [Phys. Rev. D] {10.1103/PhysRevD.104.083026}, 104, 083026

\bibitem[\protect\citeauthoryear{{Garrison-Kimmel}, {Boylan-Kolchin}, Bullock
  \& Lee}{{Garrison-Kimmel} et~al.}{2014}]{garrison-kimmel_elvis:_2014}
{Garrison-Kimmel} S.,  {Boylan-Kolchin} M.,  Bullock J.~S.,   Lee K.,  2014,
  \mn@doi [MNRAS] {10.1093/mnras/stt2377}, 438, 2578

\bibitem[\protect\citeauthoryear{{Geringer-Sameth}, Walker, Koushiappas,
  Koposov, Belokurov, Torrealba  \& Evans}{{Geringer-Sameth}
  et~al.}{2015}]{geringer-sameth_indication_2015}
{Geringer-Sameth} A.,  Walker M.~G.,  Koushiappas S.~M.,  Koposov S.~E.,
  Belokurov V.,  Torrealba G.,   Evans N.~W.,  2015, \mn@doi [Phys. Rev.]
  {10.1103/PhysRevLett.115.081101}, 115, L081101

\bibitem[\protect\citeauthoryear{Gill, Knebe  \& Gibson}{Gill
  et~al.}{2004}]{gill_evolution_2004}
Gill S. P.~D.,  Knebe A.,   Gibson B.~K.,  2004, \mn@doi [MNRAS]
  {10.1111/j.1365-2966.2004.07786.x}, 351, 399

\bibitem[\protect\citeauthoryear{Gill, Knebe  \& Gibson}{Gill
  et~al.}{2005}]{gill_evolution_2005}
Gill S. P.~D.,  Knebe A.,   Gibson B.~K.,  2005, \mn@doi [MNRAS]
  {10.1111/j.1365-2966.2004.08562.x}, 356, 1327

\bibitem[\protect\citeauthoryear{Gong et~al.,}{Gong
  et~al.}{2019}]{gong_origin_2019}
Gong C.~C.,  et~al., 2019, \mn@doi [MNRAS] {10.1093/mnras/stz1917}, 488, 3100

\bibitem[\protect\citeauthoryear{Goodenough \& Hooper}{Goodenough \&
  Hooper}{2009}]{goodenough_possible_2009}
Goodenough L.,  Hooper D.,  2009, preprint (\mn@eprint {arXiv} {0910.2998})

\bibitem[\protect\citeauthoryear{Gottl{\"o}ber, Hoffman  \&
  Yepes}{Gottl{\"o}ber et~al.}{2010}]{gottlober_constrained_2010}
Gottl{\"o}ber S.,  Hoffman Y.,   Yepes G.,  2010, preprint (\mn@eprint {arXiv}
  {1005.2687})

\bibitem[\protect\citeauthoryear{Grand \& White}{Grand \&
  White}{2021}]{grand_baryonic_2021}
Grand R. J.~J.,  White S. D.~M.,  2021, \mn@doi [MNRAS]
  {10.1093/mnras/staa3993}, 501, 3558

\bibitem[\protect\citeauthoryear{Grand et~al.,}{Grand
  et~al.}{2017}]{grand_auriga_2017}
Grand R. J.~J.,  et~al., 2017, \mn@doi [MNRAS] {10.1093/mnras/stx071}, 467, 179

\bibitem[\protect\citeauthoryear{Grand et~al.,}{Grand
  et~al.}{2019}]{grand_gas_2019}
Grand R. J.~J.,  et~al., 2019, \mn@doi [MNRAS] {10.1093/mnras/stz2928}, 490,
  4786

\bibitem[\protect\citeauthoryear{Green \& {van~den~Bosch}}{Green \&
  {van~den~Bosch}}{2019}]{green_tidal_2019}
Green S.~B.,  {van~den~Bosch} F.~C.,  2019, \mn@doi [MNRAS]
  {10.1093/mnras/stz2767}, 490, 2091

\bibitem[\protect\citeauthoryear{Green, {van~den~Bosch}  \& Jiang}{Green
  et~al.}{2021}]{green_tidal_2021}
Green S.~B.,  {van~den~Bosch} F.~C.,   Jiang F.,  2021, \mn@doi [MNRAS]
  {10.1093/mnras/stab696}, 503, 4075

\bibitem[\protect\citeauthoryear{Harris et~al.,}{Harris
  et~al.}{2020}]{harris_array_2020}
Harris C.~R.,  et~al., 2020, \mn@doi [Nature] {10.1038/s41586-020-2649-2}, 585,
  357

\bibitem[\protect\citeauthoryear{Hayashi, Navarro, Taylor, Stadel  \&
  Quinn}{Hayashi et~al.}{2003}]{hayashi_structural_2003}
Hayashi E.,  Navarro J.~F.,  Taylor J.~E.,  Stadel J.,   Quinn T.,  2003,
  \mn@doi [ApJ] {10.1086/345788}, 584, 541

\bibitem[\protect\citeauthoryear{Hoffman \& Ribak}{Hoffman \&
  Ribak}{1991}]{hoffman_constrained_1991}
Hoffman Y.,  Ribak E.,  1991, \mn@doi [AJ] {10.1086/186160}, 380, L5

\bibitem[\protect\citeauthoryear{Hunter}{Hunter}{2007}]{hunter_matplotlib_2007}
Hunter J.~D.,  2007, \mn@doi [Comput. Sci. Eng.] {10.1109/MCSE.2007.55}, 9, 90

\bibitem[\protect\citeauthoryear{Ivezi{\'c} et~al.,}{Ivezi{\'c}
  et~al.}{2019}]{ivezic_lsst_2019}
Ivezi{\'c} {\v Z}.,  et~al., 2019, \mn@doi [ApJ] {10.3847/1538-4357/ab042c},
  873, 111

\bibitem[\protect\citeauthoryear{Jones, Oliphant  \& Peterson}{Jones
  et~al.}{2011}]{jones_scipy_2011}
Jones E.,  Oliphant T.,   Peterson P.,  2011, {{SciPy Open}} Source Scientific
  Tools for {{Python}}, \url {www.scipy.org}

\bibitem[\protect\citeauthoryear{Kamionkowski, Koushiappas  \&
  Kuhlen}{Kamionkowski et~al.}{2010}]{kamionkowski_galactic_2010}
Kamionkowski M.,  Koushiappas S.~M.,   Kuhlen M.,  2010, \mn@doi [Phys. Rev. D]
  {10.1103/PhysRevD.81.043532}, 81, 043532

\bibitem[\protect\citeauthoryear{Kazantzidis, Mayer, Mastropietro, Diemand,
  Stadel  \& Moore}{Kazantzidis et~al.}{2004}]{kazantzidis_density_2004}
Kazantzidis S.,  Mayer L.,  Mastropietro C.,  Diemand J.,  Stadel J.,   Moore
  B.,  2004, \mn@doi [ApJ] {10.1086/420840}, 608, 663

\bibitem[\protect\citeauthoryear{Knebe, Libeskind, Knollmann, Yepes,
  Gottl{\"o}ber  \& Hoffman}{Knebe et~al.}{2010}]{knebe_impact_2010}
Knebe A.,  Libeskind N.~I.,  Knollmann S.~R.,  Yepes G.,  Gottl{\"o}ber S.,
  Hoffman Y.,  2010, \mn@doi [MNRAS] {10.1111/j.1365-2966.2010.16514.x}, 405,
  1119

\bibitem[\protect\citeauthoryear{Knebe, Libeskind, Knollmann,
  {Martinez-Vaquero}, Yepes, Gottl{\"o}ber  \& Hoffman}{Knebe
  et~al.}{2011a}]{knebe_luminosities_2011}
Knebe A.,  Libeskind N.~I.,  Knollmann S.~R.,  {Martinez-Vaquero} L.~A.,  Yepes
  G.,  Gottl{\"o}ber S.,   Hoffman Y.,  2011a, \mn@doi [MNRAS]
  {10.1111/j.1365-2966.2010.17924.x}, 412, 529

\bibitem[\protect\citeauthoryear{Knebe, Libeskind, Doumler, Yepes,
  Gottl{\"o}ber  \& Hoffman}{Knebe et~al.}{2011b}]{knebe_renegade_2011}
Knebe A.,  Libeskind N.~I.,  Doumler T.,  Yepes G.,  Gottl{\"o}ber S.,
  Hoffman Y.,  2011b, \mn@doi [MNRAS] {10.1111/j.1745-3933.2011.01119.x}, 417,
  L56

\bibitem[\protect\citeauthoryear{Knebe et~al.,}{Knebe
  et~al.}{2013}]{knebe_structure_2013}
Knebe A.,  et~al., 2013, \mn@doi [MNRAS] {10.1093/mnras/stt1403}, 435, 1618

\bibitem[\protect\citeauthoryear{Knollmann \& Knebe}{Knollmann \&
  Knebe}{2009}]{knollmann_ahf_2009}
Knollmann S.~R.,  Knebe A.,  2009, \mn@doi [ApJS]
  {10.1088/0067-0049/182/2/608}, 182, 608

\bibitem[\protect\citeauthoryear{Kravtsov, Gnedin  \& Klypin}{Kravtsov
  et~al.}{2004}]{kravtsov_tumultuous_2004}
Kravtsov A.~V.,  Gnedin O.~Y.,   Klypin A.~A.,  2004, \mn@doi [ApJ]
  {10.1086/421322}, 609, 482

\bibitem[\protect\citeauthoryear{Lee, Lisanti, Safdi, Slatyer  \& Xue}{Lee
  et~al.}{2016}]{lee_evidence_2016}
Lee S.~K.,  Lisanti M.,  Safdi B.~R.,  Slatyer T.~R.,   Xue W.,  2016, \mn@doi
  [Phys. Rev.] {10.1103/PhysRevLett.116.051103}, 116, L051103

\bibitem[\protect\citeauthoryear{Li, Gao, Xie  \& Guo}{Li
  et~al.}{2013}]{li_assembly_2013}
Li R.,  Gao L.,  Xie L.,   Guo Q.,  2013, \mn@doi [MNRAS]
  {10.1093/mnras/stt1551}, 435, 3592

\bibitem[\protect\citeauthoryear{Libeskind, Yepes, Knebe, Gottl{\"o}ber,
  Hoffman  \& Knollmann}{Libeskind et~al.}{2010}]{libeskind_constrained_2010}
Libeskind N.~I.,  Yepes G.,  Knebe A.,  Gottl{\"o}ber S.,  Hoffman Y.,
  Knollmann S.~R.,  2010, \mn@doi [MNRAS] {10.1111/j.1365-2966.2009.15766.x},
  401, 1889

\bibitem[\protect\citeauthoryear{Libeskind, Knebe, Hoffman, Gottl{\"o}ber,
  Yepes  \& Steinmetz}{Libeskind et~al.}{2011}]{libeskind_preferred_2011}
Libeskind N.~I.,  Knebe A.,  Hoffman Y.,  Gottl{\"o}ber S.,  Yepes G.,
  Steinmetz M.,  2011, \mn@doi [MNRAS] {10.1111/j.1365-2966.2010.17786.x}, 411,
  1525

\bibitem[\protect\citeauthoryear{Libeskind, Guo, Tempel  \& Ibata}{Libeskind
  et~al.}{2016}]{libeskind_lopsided_2016}
Libeskind N.~I.,  Guo Q.,  Tempel E.,   Ibata R.,  2016, \mn@doi [ApJ]
  {10.3847/0004-637X/830/2/121}, 830, 121

\bibitem[\protect\citeauthoryear{Libeskind et~al.,}{Libeskind
  et~al.}{2020}]{libeskind_hestia_2020}
Libeskind N.~I.,  et~al., 2020, \mn@doi [MNRAS] {10.1093/mnras/staa2541}, 498,
  2968

\bibitem[\protect\citeauthoryear{Lovell et~al.,}{Lovell
  et~al.}{2016}]{lovell_satellite_2016}
Lovell M.~R.,  et~al., 2016, \mn@doi [MNRAS] {10.1093/mnras/stw1317}, 461, 60

\bibitem[\protect\citeauthoryear{Ludlow, Navarro, Springel, Jenkins, Frenk  \&
  Helmi}{Ludlow et~al.}{2009}]{ludlow_unorthodox_2009}
Ludlow A.~D.,  Navarro J.~F.,  Springel V.,  Jenkins A.,  Frenk C.~S.,   Helmi
  A.,  2009, \mn@doi [ApJ] {10.1088/0004-637X/692/1/931}, 692, 931

\bibitem[\protect\citeauthoryear{Ludlow, Navarro, Angulo, {Boylan-Kolchin},
  Springel, Frenk  \& White}{Ludlow
  et~al.}{2014}]{ludlow_massconcentrationredshift_2014}
Ludlow A.~D.,  Navarro J.~F.,  Angulo R.~E.,  {Boylan-Kolchin} M.,  Springel
  V.,  Frenk C.,   White S. D.~M.,  2014, \mn@doi [MNRAS]
  {10.1093/mnras/stu483}, 441, 378

\bibitem[\protect\citeauthoryear{McConnachie, Irwin, Ferguson, Ibata, Lewis  \&
  Tanvir}{McConnachie et~al.}{2005}]{mcconnachie_distances_2005}
McConnachie A.~W.,  Irwin M.~J.,  Ferguson A. M.~N.,  Ibata R.~A.,  Lewis
  G.~F.,   Tanvir N.,  2005, \mn@doi [MNRAS]
  {10.1111/j.1365-2966.2004.08514.x}, 356, 979

\bibitem[\protect\citeauthoryear{McConnachie, Higgs, Thomas, Venn,
  C{\^o}t{\'e}, Battaglia  \& Lewis}{McConnachie
  et~al.}{2021}]{mcconnachie_solo_2021}
McConnachie A.~W.,  Higgs C.~R.,  Thomas G.~F.,  Venn K.~A.,  C{\^o}t{\'e} P.,
  Battaglia G.,   Lewis G.~F.,  2021, \mn@doi [MNRAS] {10.1093/mnras/staa3740},
  501, 2363

\bibitem[\protect\citeauthoryear{Molin{\'e}, {S{\'a}nchez-Conde},
  {Palomares-Ruiz}  \& Prada}{Molin{\'e}
  et~al.}{2017}]{moline_characterization_2017}
Molin{\'e} {\'A}.,  {S{\'a}nchez-Conde} M.~A.,  {Palomares-Ruiz} S.,   Prada
  F.,  2017, \mn@doi [MNRAS] {10.1093/mnras/stx026}, 466, 4974

\bibitem[\protect\citeauthoryear{Moore, Diemand  \& Stadel}{Moore
  et~al.}{2004}]{moore_age-radius_2004}
Moore B.,  Diemand J.,   Stadel J.,  2004, \mn@doi [Proceedings of the
  International Astronomical Union] {10.1017/S1743921304001127}, 2004, 513

\bibitem[\protect\citeauthoryear{Nadler et~al.,}{Nadler
  et~al.}{2021}]{nadler_constraints_2021}
Nadler E.~O.,  et~al., 2021, \mn@doi [PRL] {10.1103/PhysRevLett.126.091101},
  126, 091101

\bibitem[\protect\citeauthoryear{Navarro, Frenk  \& White}{Navarro
  et~al.}{1995}]{navarro_simulations_1995}
Navarro J.~F.,  Frenk C.~S.,   White S. D.~M.,  1995, \mn@doi [MNRAS]
  {10.1093/mnras/275.3.720}, 275, 720

\bibitem[\protect\citeauthoryear{Navarro, Frenk  \& White}{Navarro
  et~al.}{1996}]{navarro_structure_1996}
Navarro J.~F.,  Frenk C.~S.,   White S. D.~M.,  1996, \mn@doi [ApJ]
  {10.1086/177173}, 462, 563

\bibitem[\protect\citeauthoryear{Navarro, Frenk  \& White}{Navarro
  et~al.}{1997}]{navarro_universal_1997}
Navarro J.~F.,  Frenk C.~S.,   White S. D.~M.,  1997, \mn@doi [ApJ]
  {10.1086/304888}, 490, 493

\bibitem[\protect\citeauthoryear{Newton}{Newton}{2022}]{newton_hermeian_2022-1}
Newton O.,  2022, Hermeian Paper Plotting Code, Zenodo,
  \mn@doi{10.5281/zenodo.6629724}, \url {https://zenodo.org/record/6629724}

\bibitem[\protect\citeauthoryear{Newton, Cautun, Jenkins, Frenk  \&
  Helly}{Newton et~al.}{2018}]{newton_total_2018}
Newton O.,  Cautun M.,  Jenkins A.,  Frenk C.~S.,   Helly J.~C.,  2018, \mn@doi
  [MNRAS] {10.1093/mnras/sty1085}, 479, 2853

\bibitem[\protect\citeauthoryear{Newton et~al.,}{Newton
  et~al.}{2021}]{newton_constraints_2021}
Newton O.,  et~al., 2021, \mn@doi [JCAP] {10.1088/1475-7516/2021/08/062}, 2021,
  062

\bibitem[\protect\citeauthoryear{Onions et~al.,}{Onions
  et~al.}{2012}]{onions_subhaloes_2012}
Onions J.,  et~al., 2012, \mn@doi [MNRAS] {10.1111/j.1365-2966.2012.20947.x},
  423, 1200

\bibitem[\protect\citeauthoryear{Pakmor, Springel, Bauer, Mocz, Munoz, Ohlmann,
  Schaal  \& Zhu}{Pakmor et~al.}{2016}]{pakmor_improving_2016}
Pakmor R.,  Springel V.,  Bauer A.,  Mocz P.,  Munoz D.~J.,  Ohlmann S.~T.,
  Schaal K.,   Zhu C.,  2016, \mn@doi [MNRAS] {10.1093/mnras/stv2380}, 455,
  1134

\bibitem[\protect\citeauthoryear{Pawlowski}{Pawlowski}{2018}]{pawlowski_planes_2018}
Pawlowski M.~S.,  2018, \mn@doi [Modern Physics Letters A]
  {10.1142/S0217732318300045}

\bibitem[\protect\citeauthoryear{Pawlowski \& McGaugh}{Pawlowski \&
  McGaugh}{2014}]{pawlowski_perseus_2014}
Pawlowski M.~S.,  McGaugh S.~S.,  2014, \mn@doi [MNRAS] {10.1093/mnras/stu321},
  440, 908

\bibitem[\protect\citeauthoryear{Pawlowski, Ibata  \& Bullock}{Pawlowski
  et~al.}{2017}]{pawlowski_lopsidedness_2017}
Pawlowski M.~S.,  Ibata R.~A.,   Bullock J.~S.,  2017, \mn@doi [ApJ]
  {10.3847/1538-4357/aa9435}, 850, 132

\bibitem[\protect\citeauthoryear{Pe{\~n}arrubia, Navarro  \&
  McConnachie}{Pe{\~n}arrubia et~al.}{2008}]{penarrubia_tidal_2008}
Pe{\~n}arrubia J.,  Navarro J.~F.,   McConnachie A.~W.,  2008, \mn@doi [ApJ]
  {10.1086/523686}, 673, 226

\bibitem[\protect\citeauthoryear{Pe{\~n}arrubia, Benson, Walker, Gilmore,
  McConnachie  \& Mayer}{Pe{\~n}arrubia et~al.}{2010}]{penarrubia_impact_2010}
Pe{\~n}arrubia J.,  Benson A.~J.,  Walker M.~G.,  Gilmore G.,  McConnachie
  A.~W.,   Mayer L.,  2010, \mn@doi [MNRAS] {10.1111/j.1365-2966.2010.16762.x},
  406, 1290

\bibitem[\protect\citeauthoryear{Petrovi{\'c}, Serpico  \& Zaharija{\v
  s}}{Petrovi{\'c} et~al.}{2014}]{petrovic_galactic_2014}
Petrovi{\'c} J.,  Serpico P.~D.,   Zaharija{\v s} G.,  2014, \mn@doi [JCAP]
  {10.1088/1475-7516/2014/10/052}, 2014, 052

\bibitem[\protect\citeauthoryear{{Planck Collaboration} et~al.,}{{Planck
  Collaboration} et~al.}{2014}]{planck_collaboration_planck_2014}
{Planck Collaboration} et~al., 2014, \mn@doi [A\&A]
  {10.1051/0004-6361/201321591}, 571, A16

\bibitem[\protect\citeauthoryear{Poulton, Power, Robotham  \& Elahi}{Poulton
  et~al.}{2020}]{poulton_extracting_2020}
Poulton R. J.~J.,  Power C.,  Robotham A. S.~G.,   Elahi P.~J.,  2020, \mn@doi
  [MNRAS] {10.1093/mnras/stz3202}, 491, 3820

\bibitem[\protect\citeauthoryear{Prada, Klypin, Cuesta, {Betancort-Rijo}  \&
  Primack}{Prada et~al.}{2012}]{prada_halo_2012}
Prada F.,  Klypin A.~A.,  Cuesta A.~J.,  {Betancort-Rijo} J.~E.,   Primack J.,
  2012, \mn@doi [MNRAS] {10.1111/j.1365-2966.2012.21007.x}, 423, 3018

\bibitem[\protect\citeauthoryear{Profumo, Sigurdson  \& Kamionkowski}{Profumo
  et~al.}{2006}]{profumo_what_2006}
Profumo S.,  Sigurdson K.,   Kamionkowski M.,  2006, \mn@doi [Phys. Rev.]
  {10.1103/PhysRevLett.97.031301}, 97, L031301

\bibitem[\protect\citeauthoryear{Putman, Zheng, {Price-Whelan}, Grcevich,
  Johnson, Tollerud  \& Peek}{Putman et~al.}{2021}]{putman_gas_2021}
Putman M.~E.,  Zheng Y.,  {Price-Whelan} A.~M.,  Grcevich J.,  Johnson A.~C.,
  Tollerud E.,   Peek J. E.~G.,  2021, \mn@doi [ApJ]
  {10.3847/1538-4357/abe391}, 913, 53

\bibitem[\protect\citeauthoryear{Richings et~al.,}{Richings
  et~al.}{2020}]{richings_subhalo_2020}
Richings J.,  et~al., 2020, \mn@doi [MNRAS] {10.1093/mnras/stz3448}, 492, 5780

\bibitem[\protect\citeauthoryear{Riess, Fliri  \& {Valls-Gabaud}}{Riess
  et~al.}{2012}]{riess_cepheid_2012}
Riess A.~G.,  Fliri J.,   {Valls-Gabaud} D.,  2012, \mn@doi [ApJ]
  {10.1088/0004-637X/745/2/156}, 745, 156

\bibitem[\protect\citeauthoryear{Sales, Navarro, Abadi  \& Steinmetz}{Sales
  et~al.}{2007}]{sales_cosmic_2007}
Sales L.~V.,  Navarro J.~F.,  Abadi M.~G.,   Steinmetz M.,  2007, \mn@doi
  [MNRAS] {10.1111/j.1365-2966.2007.12026.x}, 379, 1475

\bibitem[\protect\citeauthoryear{{S{\'a}nchez-Conde} \&
  Prada}{{S{\'a}nchez-Conde} \& Prada}{2014}]{sanchez-conde_flattening_2014}
{S{\'a}nchez-Conde} M.~A.,  Prada F.,  2014, \mn@doi [MNRAS]
  {10.1093/mnras/stu1014}, 442, 2271

\bibitem[\protect\citeauthoryear{{S{\'a}nchez-Conde}, Cannoni, Zandanel,
  G{\'o}mez  \& Prada}{{S{\'a}nchez-Conde}
  et~al.}{2011}]{sanchez-conde_dark_2011}
{S{\'a}nchez-Conde} M.~A.,  Cannoni M.,  Zandanel F.,  G{\'o}mez M.~E.,   Prada
  F.,  2011, \mn@doi [JCAP] {10.1088/1475-7516/2011/12/011}, 2011, 011

\bibitem[\protect\citeauthoryear{Sawala et~al.,}{Sawala
  et~al.}{2016}]{sawala_apostle_2016}
Sawala T.,  et~al., 2016, \mn@doi [MNRAS] {10.1093/mnras/stw145}, 457, 1931

\bibitem[\protect\citeauthoryear{Shaya \& Tully}{Shaya \&
  Tully}{2013}]{shaya_formation_2013}
Shaya E.~J.,  Tully R.~B.,  2013, \mn@doi [MNRAS] {10.1093/mnras/stt1714}, 436,
  2096

\bibitem[\protect\citeauthoryear{Simpson, Grand, G{\'o}mez, Marinacci, Pakmor,
  Springel, Campbell  \& Frenk}{Simpson et~al.}{2018}]{simpson_quenching_2018}
Simpson C.~M.,  Grand R. J.~J.,  G{\'o}mez F.~A.,  Marinacci F.,  Pakmor R.,
  Springel V.,  Campbell D. J.~R.,   Frenk C.~S.,  2018, \mn@doi [MNRAS]
  {10.1093/mnras/sty774}, 478, 548

\bibitem[\protect\citeauthoryear{Sorce et~al.,}{Sorce
  et~al.}{2016}]{sorce_cosmicflows_2016}
Sorce J.~G.,  et~al., 2016, \mn@doi [MNRAS] {10.1093/mnras/stv2407}, 455, 2078

\bibitem[\protect\citeauthoryear{Springel}{Springel}{2010}]{springel_e_2010}
Springel V.,  2010, \mn@doi [MNRAS] {10.1111/j.1365-2966.2009.15715.x}, 401,
  791

\bibitem[\protect\citeauthoryear{Springel et~al.,}{Springel
  et~al.}{2008}]{springel_aquarius_2008}
Springel V.,  et~al., 2008, \mn@doi [MNRAS] {10.1111/j.1365-2966.2008.14066.x},
  391, 1685

\bibitem[\protect\citeauthoryear{Srisawat et~al.,}{Srisawat
  et~al.}{2013}]{srisawat_sussing_2013}
Srisawat C.,  et~al., 2013, \mn@doi [MNRAS] {10.1093/mnras/stt1545}, 436, 150

\bibitem[\protect\citeauthoryear{Teyssier, Johnston  \& Kuhlen}{Teyssier
  et~al.}{2012}]{teyssier_identifying_2012}
Teyssier M.,  Johnston K.~V.,   Kuhlen M.,  2012, \mn@doi [MNRAS]
  {10.1111/j.1365-2966.2012.21793.x}, 426, 1808

\bibitem[\protect\citeauthoryear{{The Astropy Collaboration} et~al.,}{{The
  Astropy Collaboration} et~al.}{2013}]{the_astropy_collaboration_astropy_2013}
{The Astropy Collaboration} et~al., 2013, \mn@doi [A\&A]
  {10.1051/0004-6361/201322068}, 558, A33

\bibitem[\protect\citeauthoryear{{The Astropy Collaboration} et~al.,}{{The
  Astropy Collaboration} et~al.}{2018}]{the_astropy_collaboration_astropy_2018}
{The Astropy Collaboration} et~al., 2018, \mn@doi [AJ]
  {10.3847/1538-3881/aabc4f}, 156, 123

\bibitem[\protect\citeauthoryear{{The CTA Consortium}}{{The CTA
  Consortium}}{2019}]{the_cta_consortium_science_2019}
{The CTA Consortium} 2019, Science with the {{Cherenkov Telescope Array}}.
{WORLD SCIENTIFIC}, \mn@doi{10.1142/10986}, \url
  {https://www.worldscientific.com/worldscibooks/10.1142/10986}

\bibitem[\protect\citeauthoryear{Tully et~al.,}{Tully
  et~al.}{2013}]{tully_cosmicflows-2_2013}
Tully R.~B.,  et~al., 2013, \mn@doi [AJ] {10.1088/0004-6256/146/4/86}, 146, 86

\bibitem[\protect\citeauthoryear{Van~Rossum \& Drake}{Van~Rossum \&
  Drake}{2009}]{van_rossum_python_2009}
Van~Rossum G.,  Drake F.~L.,  2009, Python 3 {{Reference Manual}}.
{CreateSpace}, {Scotts Valley, CA}

\bibitem[\protect\citeauthoryear{Virtanen et~al.,}{Virtanen
  et~al.}{2020}]{virtanen_scipy_2020}
Virtanen P.,  et~al., 2020, \mn@doi [Nature Methods]
  {10.1038/s41592-019-0686-2}, 17, 261

\bibitem[\protect\citeauthoryear{Vogelsberger, Marinacci, Torrey  \&
  Puchwein}{Vogelsberger et~al.}{2020}]{vogelsberger_cosmological_2020}
Vogelsberger M.,  Marinacci F.,  Torrey P.,   Puchwein E.,  2020, \mn@doi
  [Nature Reviews Physics] {10.1038/s42254-019-0127-2}, 2, 42

\bibitem[\protect\citeauthoryear{Wan, Oliver, Lewis, Read  \& Collins}{Wan
  et~al.}{2020a}]{wan_origin_2020}
Wan Z.,  Oliver W.~H.,  Lewis G.~F.,  Read J.~I.,   Collins M. L.~M.,  2020a,
  \mn@doi [MNRAS] {10.1093/mnras/stz3477}, 492, 456

\bibitem[\protect\citeauthoryear{Wan et~al.,}{Wan
  et~al.}{2020b}]{wan_tidal_2020}
Wan Z.,  et~al., 2020b, \mn@doi [Nature] {10.1038/s41586-020-2483-6}, 583, 768

\bibitem[\protect\citeauthoryear{Wang, Bose, Frenk, Gao, Jenkins, Springel  \&
  White}{Wang et~al.}{2020}]{wang_universal_2020}
Wang J.,  Bose S.,  Frenk C.~S.,  Gao L.,  Jenkins A.,  Springel V.,   White S.
  D.~M.,  2020, \mn@doi [Nature] {10.1038/s41586-020-2642-9}, 585, 39

\bibitem[\protect\citeauthoryear{Warnick, Knebe  \& Power}{Warnick
  et~al.}{2008}]{warnick_tidal_2008}
Warnick K.,  Knebe A.,   Power C.,  2008, \mn@doi [MNRAS]
  {10.1111/j.1365-2966.2008.12992.x}, 385, 1859

\bibitem[\protect\citeauthoryear{Weinberger, Springel  \& Pakmor}{Weinberger
  et~al.}{2020}]{weinberger_arepo_2020}
Weinberger R.,  Springel V.,   Pakmor R.,  2020, \mn@doi [ApJS]
  {10.3847/1538-4365/ab908c}, 248, 32

\bibitem[\protect\citeauthoryear{Wetzel, Deason  \& {Garrison-Kimmel}}{Wetzel
  et~al.}{2015}]{wetzel_satellite_2015}
Wetzel A.~R.,  Deason A.~J.,   {Garrison-Kimmel} S.,  2015, \mn@doi [ApJ]
  {10.1088/0004-637X/807/1/49}, 807, 49

\bibitem[\protect\citeauthoryear{Xing, Zhao, Aoki, Honda, Li, Ishigaki  \&
  Matsuno}{Xing et~al.}{2019}]{xing_evidence_2019}
Xing Q.-F.,  Zhao G.,  Aoki W.,  Honda S.,  Li H.-N.,  Ishigaki M.~N.,
  Matsuno T.,  2019, \mn@doi [Nature Astronomy] {10.1038/s41550-019-0764-5}, 3,
  631

\bibitem[\protect\citeauthoryear{Yepes, Gottl{\"o}ber  \& Hoffman}{Yepes
  et~al.}{2014}]{yepes_dark_2014}
Yepes G.,  Gottl{\"o}ber S.,   Hoffman Y.,  2014, \mn@doi [New Astronomy
  Reviews] {10.1016/j.newar.2013.11.001}, 58, 1

\bibitem[\protect\citeauthoryear{Zavala \& Frenk}{Zavala \&
  Frenk}{2019}]{zavala_dark_2019}
Zavala J.,  Frenk C.~S.,  2019, \mn@doi [Galaxies] {10.3390/galaxies7040081},
  7, 81

\bibitem[\protect\citeauthoryear{{van den Bosch}}{{van den
  Bosch}}{2017}]{van_den_bosch_dissecting_2017}
{van den Bosch} F.~C.,  2017, \mn@doi [MNRAS] {10.1093/mnras/stx520}, 468, 885

\bibitem[\protect\citeauthoryear{{van den Bosch} \& Ogiya}{{van den Bosch} \&
  Ogiya}{2018}]{van_den_bosch_dark_2018}
{van den Bosch} F.~C.,  Ogiya G.,  2018, \mn@doi [MNRAS]
  {10.1093/mnras/sty084}, 475, 4066

\bibitem[\protect\citeauthoryear{van~der Walt, Colbert  \& Varoquaux}{van~der
  Walt et~al.}{2011}]{walt_numpy_2011}
van~der Walt S.,  Colbert S.~C.,   Varoquaux G.,  2011, \mn@doi [Comput. Sci.
  Eng.] {10.1109/MCSE.2011.37}, 13, 22

\makeatother
\end{thebibliography}



\appendix
\section{A comparison of the \textit{J}-factors of all field halo populations}
\label{sec:Appendix:17_11_Jfactors}
\begin{figure*}%
    \centering%
	\includegraphics[width=\textwidth]{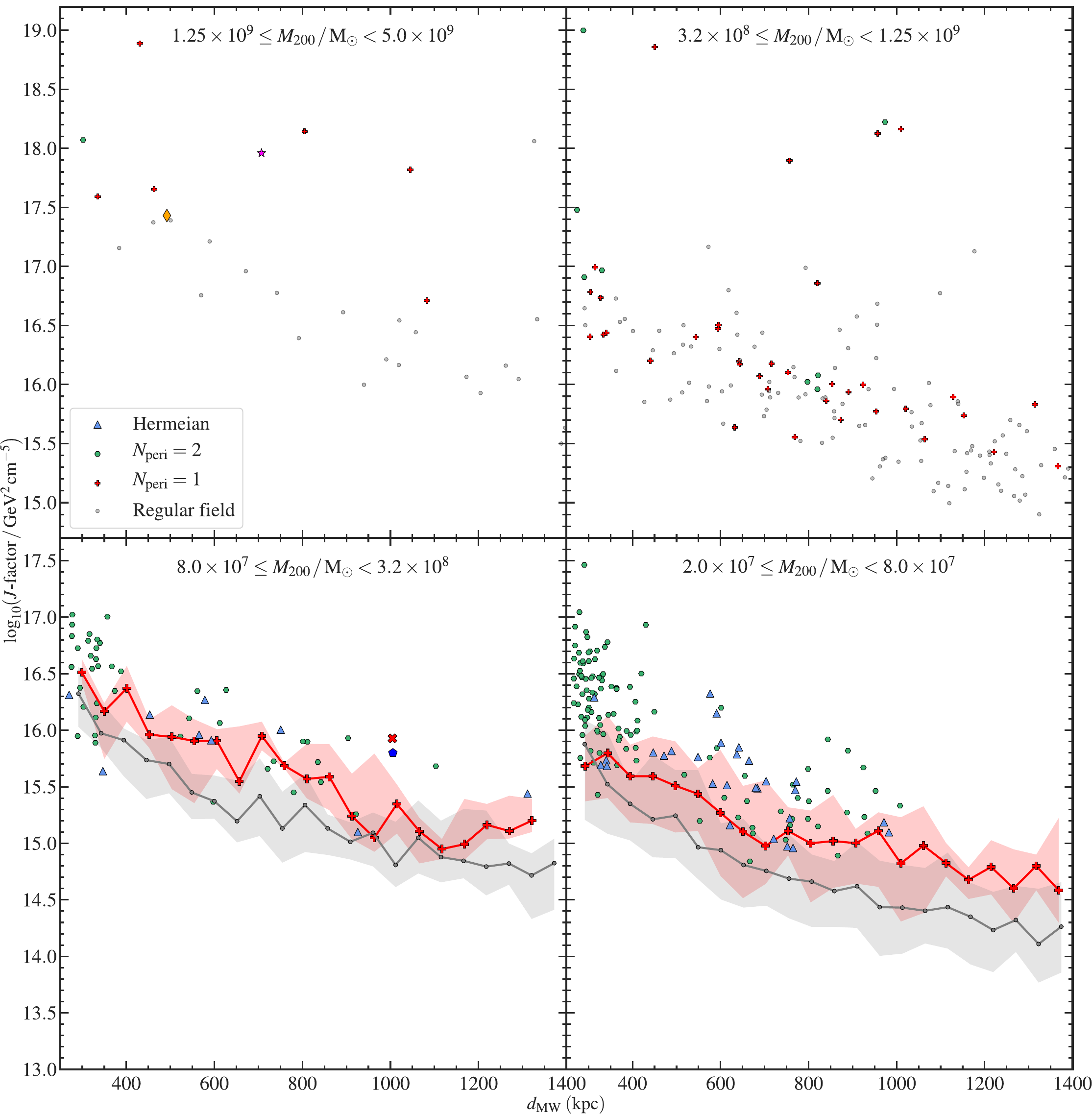}%
	\caption{The \Jfactor{}s of haloes in the \code{17_11} simulation as a function of their distances from the Milky Way analogue. We plot the \hermeian{}, backsplash, and regular field populations in four bins in halo mass with approximately equal widths in log-space. As in \figref{fig:Methods:hermeian_distance}, we plot the four \hermeian{} galaxies with their own distinguishing markers. In the two lowest-mass bins, where the sizes of the \Nperi[1] and regular field halo populations are large, we plot the medians (solid lines) and \percent{68} scatters (shaded regions) for each population. The edge of the lowest mass bin, $\Mvir{}=\Msun[{2\times 10^7}]$, is equal to the mass of $100$~DM particles in the high-resolution \Hestia{} simulations.
	}%
	\label{fig:App:17_11_Jfactors}%
	\vspace{-10pt}%
\end{figure*}%

Any object containing DM could emit radiation from the annihilation of DM particles. The detectability of this signal can be quantified using the \Jfactor{}, which depends only on the spatial distribution of the DM (see \eqnref{eq:Results:DM_searches:J_factor_NFW}). This facilitates a simple comparison of different classes of astrophysical object for their suitability as targets to detect an annihilation signal.
In \figref{fig:App:17_11_Jfactors}, we plot the \Jfactor{}s of the \hermeian{}, \Nperi[1] and \Nperi[2] backsplash, and regular field halo populations in the \code{17_11} simulation as a function of distance from the Milky Way analogue. The \Jfactor{} depends most sensitively on the distance of the object from the observer and the concentration of its DM component, and is only linearly dependent on the halo mass. However, the steep mass function of the field haloes produces a larger dynamic range in halo mass than that of the Milky~{Way--field} halo distance and the halo concentration. This affects the \Jfactor{} significantly. Therefore, to compare each halo population we divide the samples into four bins in halo mass. \figref{fig:App:17_11_Jfactors} shows that the \hermeian{} and \Nperi[2] backsplash haloes have higher \Jfactor{}s, on average, than regular field haloes because they have higher concentrations (see \figref{fig:Results:c200_vs_vmax_vmax_vs_m200}). In this simulation, the \Nperi[2] backsplash haloes are typically found close to \Rvir{} of the Milky Way analogue. They therefore have high \Jfactor{}s compared to other halo populations in the same mass bin, which makes them interesting targets as sources of DM annihilation signals.


\bsp	
\label{lastpage}
\end{document}